\documentclass[fleqn,usenatbib]{mnras}

\usepackage[T1]{fontenc}
\usepackage{ae,aecompl}

\usepackage{graphicx}	
\usepackage{amsmath}	
\usepackage{amsfonts}   
\usepackage{amssymb}	
\usepackage{gensymb}    
\usepackage{soul} 
\usepackage{pdflscape}  

\def \apjs {ApJS}
\def \apjl {ApJL}
\def \aap {A\&A}

\newcommand{\ergcms} {erg cm$^{-2}$ s$^{-1}$}
\newcommand{\ergs} {erg s$^{-1}$}
\newcommand{\msol} {M$_{\odot}$}

\def\lesssim{\mathrel{\hbox{\rlap{\hbox{\lower4pt\hbox{$\sim$}}}\hbox{$<$}}}}
\def\gtrsim{\mathrel{\hbox{\rlap{\hbox{\lower4pt\hbox{$\sim$}}}\hbox{$>$}}}}
\newcommand{\ang} {\r{A}$\,$}
\newcommand{\halpha} {$\mathrm{H\alpha}$}
\newcommand{\hbeta} {$\mathrm{H\beta}\,$}
\newcommand{\hgamma} {$\mathrm{H\gamma}\,$}
\newcommand{\sfr} {\msol yr$^{-1}$}
\newcommand{\sfi} {\msol yr$^{-1}$ kpc$^{-2}$}
\newcommand{\SFI} {$\Sigma_{\text{SFR}}$}
\long\def\symbolfootnote[#1]#2{\begingroup%
\def\thefootnote{\fnsymbol{footnote}}\footnote[#1]{#2}\endgroup}

\title[WR population of IC10]{Revealing the nebular properties and Wolf-Rayet population of IC10 with Gemini/GMOS}

\author[K. Tehrani et al.]{
Katie Tehrani,$^{1}$\thanks{k.tehrani@sheffield.ac.uk}
Paul A. Crowther,$^{1}$
I. Archer$^{1}$
\\
$^{1}$Department of Physics and Astronomy, University of Sheffield, Sheffield, S3 7RH, UK
}

\date{Accepted XXX. Received YYY; in original form ZZZ}

\pubyear{2016}

\begin{document}
\label{firstpage}
\pagerange{\pageref{firstpage}--\pageref{lastpage}}
\maketitle

\begin{abstract}
  We present a deep imaging and spectroscopic survey of the Local Group irregular galaxy IC10 using Gemini North and GMOS to unveil its global Wolf-Rayet (WR) population. We obtain a star formation rate (SFR) of 0.045 $\pm$ 0.023 \sfr, for IC10 from the nebular \halpha\ luminosity, which is comparable to the SMC. We also present a revised nebular oxygen abundance of log(O/H) + 12 = 8.40 $\pm$ 0.04, comparable to the LMC. It has previously been suggested that for IC10 to follow the WR subtype-metallicity dependance seen in other Local Group galaxies, a large WN population awaits discovery. Our search revealed 3 new WN stars, and 6 candidates awaiting confirmation, providing little evidence to support this claim. The new global WR star total of 29 stars is consistent with the LMC population when scaled to the reduced SFR of IC10. For spectroscopically confirmed WR stars, the WC/WN ratio is lowered to 1.0, however including all potential candidates, and assuming those unconfirmed to be WN stars, would reduce the ratio to $\sim$0.7. We attribute the high WC/WN ratio to the high star formation surface density of IC10 relative to the Magellanic Clouds, which enhances the frequency of high mass stars capable of producing WC stars.

\end{abstract}





\section{Introduction}


IC10 is a barred irregular galaxy in the Local Group, considered by some as a blue compact dwarf \citep{ric2001}. Situated beyond the plane of the Milky Way ($b$=$-$3.3$\degree$), attempts to study this galaxy are hindered by a high galactic foreground reddening of E(B-V) = 0.77 \citep{ric2001}. With recent distance determinations ranging from 660 kpc \citep{gon2012} to 817 kpc \citep{san2008}, we adopt an IC10 distance of 740 $\pm$ 20 kpc\footnote{Taken from an average of four different methods; RR Lyrae 820 $\pm$ 80 kpc \citep{san2008}, PNLF 660 $\pm$ 25 kpc \citep{gon2012}, tip of the Red Giant Branch 740 $\pm$ 60 kpc and Cepheids 740 $\pm$ 60 kpc \citep{tul2013}.}. \citet{mcc2012} derived a stellar mass of 7.5 $\times$ 10$^{7}$ \msol\ for IC10, an order of magnitude lower than the Small Magellanic Cloud (SMC). IC10 is gas rich with an atomic hydrogen content of 4.4 $\times$ 10$^{7}$ \msol, adjusted to our adopted distance. The galaxy is also metal-poor, with an oxygen abundance measurement of log(O/H) + 12  = 8.26 \citep{gar1990}. This metallicity is intermediate between those of the Magellanic Clouds, making them good comparative galaxies.

The recent star formation history of IC10 is very uncertain. Studies of the neutral hydrogen content by \citet{wil1998} revealed that the gas distribution has been shaped by stellar winds rather than supernovae explosions, suggesting the observed interstellar medium features are relatively young. From this it was suggested that the galaxy is currently undergoing a starburst episode which began approximately 10 million years ago. The current star formation rate was found to be 0.07 \sfr by \citet{gre1996}, which again is between those of the Magellanic Clouds \citep{ken2008}. The star formation intensity however, is much greater than both at 0.049 \msol yr$^{-1}$ kpc$^{-2}$ \citep{cro2009}, due to the small physical size of IC10. 

Wolf Rayet (WR) stars are post main sequence massive stars progressing through the helium-burning stages of stellar evolution. These stars are dominated by strong winds and high mass loss rates during which they are stripped of their hydrogen, and sometimes helium, envelopes. This distinction lends itself nicely to the criteria for two WR spectral types, the nitrogen sequence (WN) and the carbon sequence (WC), both of which can be divided into further ionization subclasses. 
The two spectral types are thought to be linked through an evolutionary chain known as the Conti scenario which has since been adapted by \citet{cro2007} to take into consideration initial stellar mass. 

Narrow-band photometry is useful for identifying potential WR candidates, because the strong winds associated with these stars manifest as broad emission line features within the stellar spectrum. Matching narrow-band filters with the emission line wavelengths helps to reveal these stars, especially in crowded environments. This method alone however, is not sufficient to confirm a WR star, or to classify a spectral type. Instead confirmation requires spectroscopy in order to calculate the relevant emission line ratios outlined in \citet{smi1996} for WN stars, and in \citet{cro1998} for WC stars.

Previous surveys of IC10 have been successful at finding WR stars within this galaxy despite the high foreground extinction.  \citet{mas1992} first began the search after suspecting a large population of massive stars would be likely when considering the number of H\,{\sc ii} regions identified within the galaxy \citep{hod1990}. Further studies such as \citet{mas1995, roy2001, mas2002a, cro2003} led to the confirmation of 26 WR stars, hereafter referred to as M\#, and R\# depending on which collaboration initially identified the candidate. The discovery of these WR stars was curious. Not only does IC10 now have the highest surface density of WR stars in the Local Group, the ratio of WC/WN spectral types does not agree with that expected from evolution models for a galaxy of such a low metallicity.

Currently the WC/WN ratio stands at 1.3, which is an order of magnitude higher than other metal poor star forming galaxies such as the LMC (0.2) and SMC (0.1) \citep{bre1999, neu2012a, mas2014, mas2015a, foe2003a}. It has been proposed that either an unusual starburst has occurred, or there are further WN stars residing unnoticed within IC10 \citep{mas2002a}. Therefore, to confidently verify the WR content of IC10 we must first be satisfied the search is complete, especially in the context of the recent discovery of unusually faint WN stars in the LMC \citep{mas2014, neu2017}.

The purpose of our study therefore, is to use deep narrow-band imaging to establish if a hidden population of WR stars is a plausible explanation for the apparently abnormal WC/WN ratio. Since the metallicity of IC10 is intermediate between the LMC and SMC, we also look to compare the properties of WR stars found in each galaxy. Finally we aim to identify the fraction of IC10 WR stars residing in binary systems. 
The new photometric and spectroscopic data to achieve these aims are presented in Sect.~\ref{sec:obs}. Sect.~\ref{sec:neb_res} focuses on the results from nebular emission including an update of the metallicity and star formation rate (SFR). In Sect.~\ref{sec:res} we present the stellar results from these observations, a discussion of these results in Sect.~\ref{sec:dis}, and brief conclusions in Sect.~\ref{sec:con}.


\section{Observations.}
\label{sec:obs}


\subsection{Imaging}
\label{sec:photometry}

\begin{figure*}
	\includegraphics[width=\textwidth]{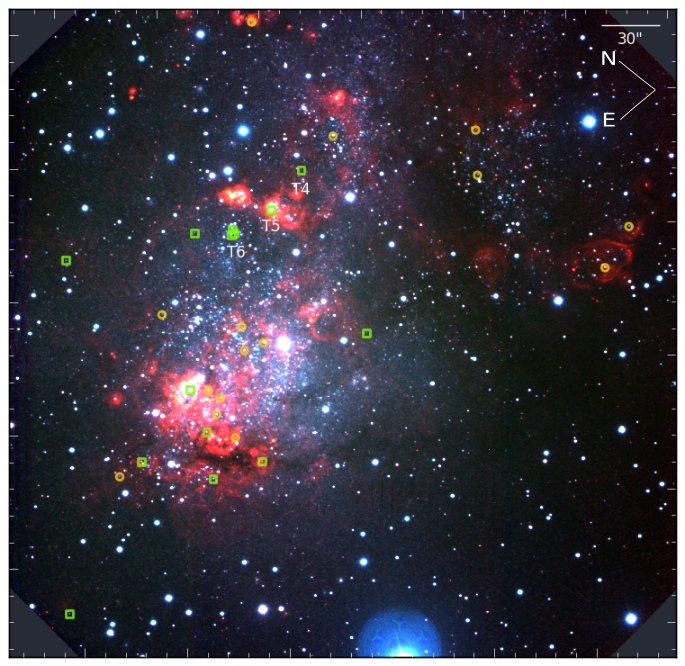}
        \caption{Gemini GMOS colour composite image of IC10 showing the relative positions of previously confirmed WN stars (green squares) and WC stars (yellow circles), with the 3 new WN stars labelled. Field of view shown in image is 300\arcsec\ $\times$ 314\arcsec, corresponding to 1.1 $\times$ 1.1 kpc at a distance of 740 kpc. RGB image generated from red-\halpha ($\lambda$656nm), green-g($\lambda$475nm), and blue-He\,{\sc ii}($\lambda$468nm) filter images.}
    \label{fig:allWR}
\end{figure*}

We obtained deep imaging observations of IC10 on the 24 September 2009 using the GMOS instrument \citep{hoo2004} on the 8m Gemini North telescope at Mauna Kea (ID GN-2009B-Q-9, PI Crowther). Due to the emission line nature of WR stars, narrow-band filters are more suited to identify potential candidates, therefore four narrow-band filters and one broadband filter were selected, with details outlined in Table~\ref{tab:gmosfilters}. Two of the narrow-band filters were centred on strong emission lines, and the remaining two were continuum filters, denoted by the suffix C. The He\,{\sc ii} 4686 emission line is particularly strong within all WR subtypes, therefore to identify He\,{\sc ii} excess candidates both continuum and emission-line imaging is necessary.
The 330$\arcsec$ $\times$ 330$\arcsec$ field of view on GMOS-N is composed of three CCDs separated by a 2.8$\arcsec$ gap. To compensate for this gap, IC10 was observed three times through each filter, with each subsequent image shifted by 5$\arcsec$. The resultant field of view can be seen in the colour composite image shown in Fig.~\ref{fig:allWR}, where the positions of all confirmed WR stars have also been highlighted.

\begin{table}
\caption{Characteristics of the five filters used with GMOS and the imaging quality of IC10}
\label{tab:gmosfilters}
\begin{tabular}{c@{\hspace{3mm}}c@{\hspace{3mm}}c@{\hspace{3mm}}c@{\hspace{3mm}}c@{\hspace{3mm}}}
\hline
\hline
Filter & $\lambda_{c}$ & FWHM & T$_{exp}$ & FWHM \\
 & [nm] & [nm] & [s] & [$\arcsec$] \\
\hline
He\,{\sc ii}      & 468 & 8   & 3 x 1750 & 0.59 \\
He\,{\sc ii}C     & 478 & 8   & 3 x 1750 & 0.56 \\
\halpha   & 656 & 7   & 3 x 60   & 0.53 \\
\halpha C & 662 & 6   & 3 x 60   & 0.52 \\
g         & 475 & 154 & 3 x 30   & 0.59 \\
\hline
\hline
\end{tabular}
\end{table}

\begin{figure}
	\includegraphics[width=\columnwidth]{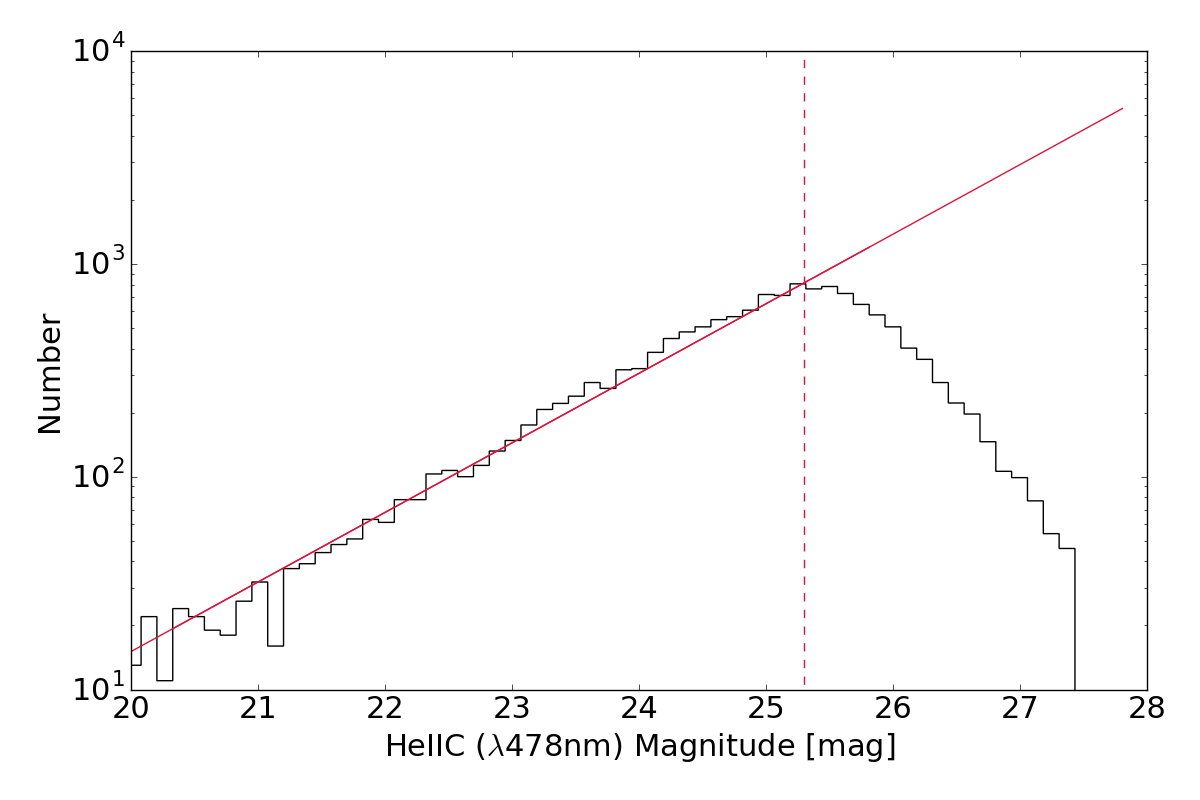}
    \caption{Logarithmic histogram showing the He\,{\sc ii}C($\lambda$478nm) mag for all sources within the field of view. The peak of the curve indicates this survey is complete to an apparent He\,{\sc ii}C magnitude of 25.3, or absolute magnitude of -2.4 for an adopted E(B-V)=0.917, A$_{4780}$=1.194 A$_{V}$=3.50 and distance modulus of 24.3 mag.}
    \label{fig:magerr}
\end{figure}

\begin{figure*}
  \includegraphics[width=\textwidth]{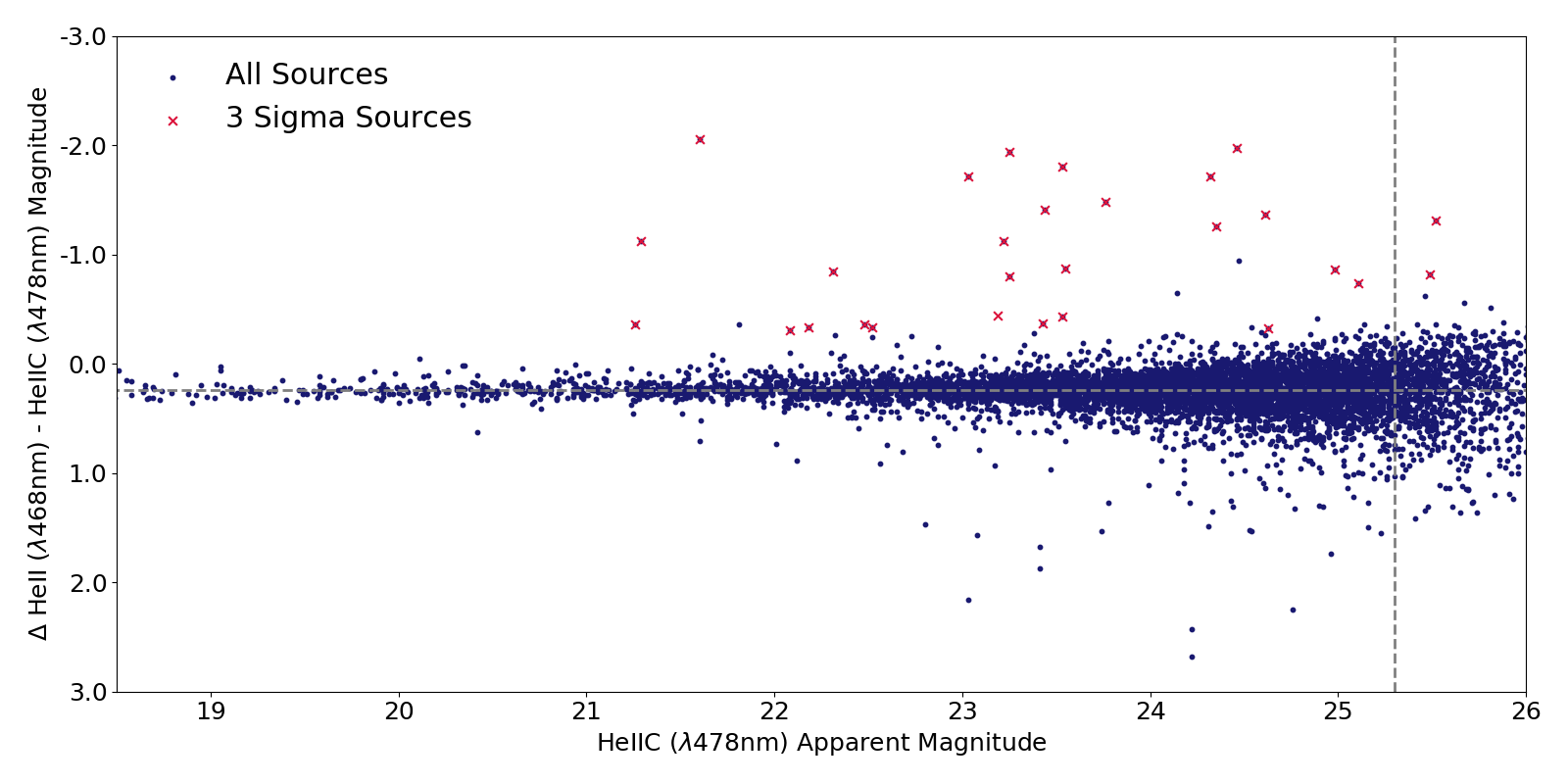}
  \includegraphics[width=\textwidth]{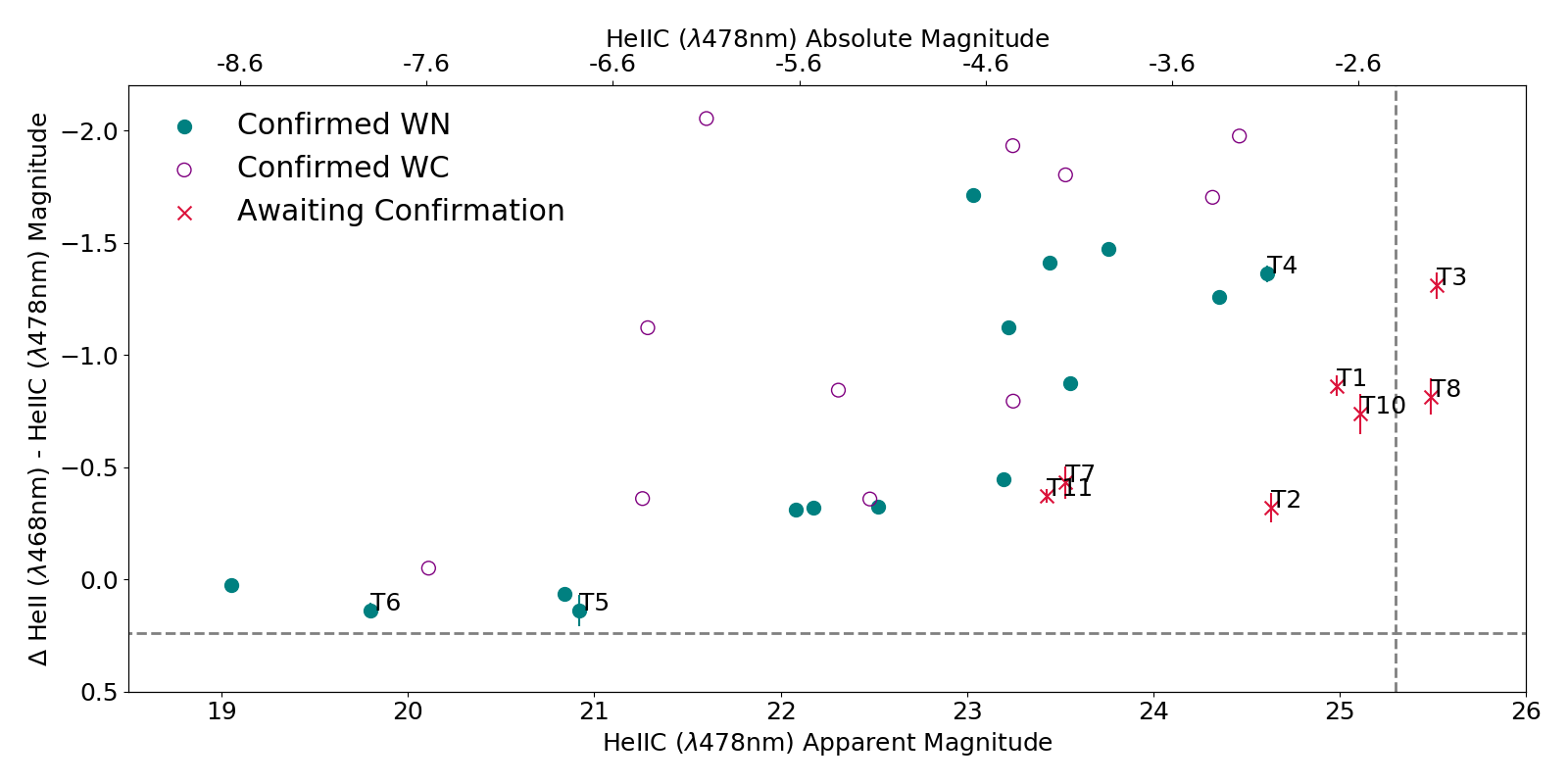}  
  \caption{Upper panel shows the relationship between continuum magnitude and He\,{\sc ii} 4686 excess magnitude for all sources identified by DAOPHOT in IRAF. The He\,{\sc ii} emission zero-point lies at -0.24, as shown by the horizontal dashed line, due to the high extinction towards IC10. The vertical dashed line corresponds to the survey limit. The majority lie below the zero-excess line indicating no He\,{\sc ii} emission line excess, and a considerable proportion of the data points are also below the detection limits of this survey, therefore too faint to be considered robust. The red crosses correspond to sources which were found to have a He\,{\sc ii} detection greater than 3$\sigma$. The lower panel is restricted to confirmed WR candidates (WN - closed green circles, WC - open purple circles) and those awaiting confirmation (red crosses). Note that the majority of unconfirmed candidates are much fainter than others in the sample, and there are no candidates included which exceed our He\,{\sc ii}C survey limit of 25.3 mag. Both apparent magnitude and absolute magnitudes axis are shown, with absolute magnitudes derived from A$_{4780}$=3.39 mag and a distance modulus of 24.3 mag.}
  \label{fig:deltamagAwaitingConf}
\end{figure*}

Images in the same filter were aligned and combined in IRAF before using the DAOPHOT package to find the relative magnitudes of all sources with an appropriate point spread function \citep{ste1987}. To convert narrow-band values to apparent magnitudes a zero point correction was applied, derived from observations of the standard star BD +28 4211.

We also obtained broad g-band imaging which was reduced using a similar method, however standard star observations were unavailable. Instead, isolated sources were selected from within the GMOS field of view for which apparent magnitudes were obtained from PanSTARRS1 g-band observations \citep{cha2016}. These standard stars provided an average g-band zero point correction for the remainder of the GMOS data.
 
Of the 37 previous WR candidate stars, 36 were located and identified within our field of view. The exception, M22, had previously been dismissed as a potential WR by \citet{mas1995}, and of the 37 candidates, 26 have been confirmed as WR stars. Photometric results for confirmed WR stars are shown in Table~\ref{tab:photdata} where positions have been astrometrically corrected based on PanSTARRS data. New WR candidates were selected from a combination of image subtraction techniques, blinking of the He\,{\sc ii} and He\,{\sc ii}C images, and quantitative relative magnitude comparisons to reveal an excess of He\,{\sc ii} 4686 emission. A visual inspection of all HeII excess sources with a greater than 3 sigma detection was completed to remove spurious sources, and 11 new emission candidates were revealed with their photometric properties outlined in Table~\ref{tab:potentialWR} in the appendix. We note that T8 appears extended with FWHM measurements of 0.64\arcsec $\times$ 0.98\arcsec, compared to the average FWHM of 0.66\arcsec $\times$ 0.63\arcsec for point sources. We therefore advise that although this source exhibits a HeII excess it is unlikely to be confirmed a WR star. Fig.~\ref{fig:deltamagAwaitingConf} shows the He\,{\sc ii} excesses as a function of continuum magnitude for the potential WR candidates, along with a comparison including all DAOPHOT sources.
 
Fig.~\ref{fig:magerr} shows the logarithmic number count distribution of He\,{\sc ii}C magnitudes across all sources located by DAOPHOT within our field of view with a least squares polynomial fit applied to the region in which this relationship is linear. The point at which this relationship breaks down, and the slope turns over indicates the faintest magnitude we consider this survey to be complete to within a 3$\sigma$ error, which for He\,{\sc ii}C corresponds to an apparent magnitude of 25.3 mag. Using the same principle for the He\,{\sc ii} photometry results in a faint magnitude limit of 24.7 mag. From our compiled list of 11 WR candidates, two fall outside this completeness limit (T3 and T8).

The He\,{\sc ii}C apparent magnitude was corrected for interstellar extinction using an average E(B-V) of 0.92, discussed in Sect.~\ref{sec:ebv}, so A$_{4780}$=3.4 mag when adopting A$_{4780}$/A$_{V}$=1.194 from the Galactic extinction law \citep{sea1979}. We adopt a distance modulus of 24.3 $\pm$ 0.07 mag, corresponding to a distance of 740 $\pm$ 20 kpc, and from that we reach an absolute magnitude survey limit of -2.4 mag for the He\,{\sc ii} continuum.

\begin{landscape}
\begin{table}
\begin{center}
  \caption{Photometric magnitudes taken from GMOS narrow-band filter images of all currently confirmed WR stars residing within IC10, listed in increasing RA order. HL\# refers to the H\,{\sc ii} region each star is associated with, outlined in \citet{hod1990}. Old spectral types are taken from \citet{mas1995, roy2001, mas2002a, cro2003}. New spectral classes have been determined from GMOS spectroscopic data (discussed in Sect.~\ref{sec:confirmedWR}). Interstellar extinction magnitudes were derived using method outlined in Sect.~\ref{sec:ebv} using both photometric data shown here and spectral classifications. M$_{v}$ absolute v-band magnitudes are derived from spectroscopically measured magnitudes and an adopted distance of 0.74Mpc.}
\label{tab:photdata}
\begin{tabular}{c@{\hspace{1.5mm}}c@{\hspace{1.5mm}}c@{\hspace{1.5mm}}c@{\hspace{1.5mm}}c@{\hspace{1.5mm}}c@{\hspace{1.5mm}}c@{\hspace{1.5mm}}c@{\hspace{1.5mm}}c@{\hspace{1.5mm}}
c@{\hspace{1.5mm}}c@{\hspace{1.5mm}}c@{\hspace{1.5mm}}c@{\hspace{0.5mm}}c@{\hspace{0.5mm}}}
\hline
\hline
ID & HL & RA & Dec & Old Spectral & ref & New Spectral& He\,{\sc ii} & He\,{\sc ii}C & \halpha & \halpha C & g & E(B-V) & M$_{v}$ \\
 & & \multicolumn{2}{c}{[J2000]} & Type & & Type & [mag] & [mag] & [mag] & [mag] & [mag] & [mag] & [mag] \\
\hline
M1 & 2/3         & 00:19:56.96 & 59:17:07.6 & WC4-5  & c & WC4-5       & 21.624 $\pm$ 0.024 &                    & 21.028 $\pm$ 0.047 & 21.597 $\pm$ 0.019 & 21.560 $\pm$ 0.037 & \textit{0.92 $\pm$ 0.26} & -5.62 $\pm$ 0.87 \\
M2 & 6           & 00:19:59.62 & 59:16:54.7 & WC4    & c & WC4         & 21.235 $\pm$ 0.020 &                    & 19.908 $\pm$ 0.053 & 20.764 $\pm$ 0.050 &                    & \textit{0.92 $\pm$ 0.26} & -5.86 $\pm$ 0.87 \\
R6 & 10          & 00:20:02.99 & 59:18:26.9 & WC4    & c & WC4         & 22.453 $\pm$ 0.014 & 23.248 $\pm$ 0.020 & 20.723 $\pm$ 0.033 & 21.690 $\pm$ 0.041 & 22.441 $\pm$ 0.022 & 1.28 $\pm$ 0.04          & -5.75 $\pm$ 0.14 \\
R5 &             & 00:20:04.27 & 59:18:06.2 & WC4-5  & c & WC4-5       & 22.120 $\pm$ 0.027 & 22.479 $\pm$ 0.026 & 20.951 $\pm$ 0.035 & 21.346 $\pm$ 0.036 & 21.822 $\pm$ 0.024 & 0.94 $\pm$ 0.04          & -5.36 $\pm$ 0.14 \\
M4 &             & 00:20:11.55 & 59:18:57.6 & WC4-5  & c & WC4-5       & 20.060 $\pm$ 0.007 & 20.111 $\pm$ 0.005 & 18.718 $\pm$ 0.007 & 19.169 $\pm$ 0.011 & 19.510 $\pm$ 0.006 & 0.91 $\pm$ 0.01          & -7.67 $\pm$ 0.07 \\
M5 & 29          & 00:20:12.83 & 59:20:08.0 & WNE/C4 & c & WNE/C4      & 21.323 $\pm$ 0.008 & 23.035 $\pm$ 0.008 & 19.665 $\pm$ 0.062 & 21.783 $\pm$ 0.032 & 21.804 $\pm$ 0.027 & 1.04 $\pm$ 0.03          & -4.86 $\pm$ 0.11 \\
T4 &             & 00:20:14.47 & 59:18:49.9 &        &   & WNE         & 23.249 $\pm$ 0.016 & 24.612 $\pm$ 0.034 & 22.765 $\pm$ 0.123 & 23.377 $\pm$ 0.121 & 23.958 $\pm$ 0.093 & 1.03 $\pm$ 0.10          & -3.48 $\pm$ 0.35 \\
R13&             & 00:20:15.62 & 59:17:21.4 & WN5    & c & WN5         & 23.092 $\pm$ 0.018 & 24.351 $\pm$ 0.025 & 22.238 $\pm$ 0.063 & 23.306 $\pm$ 0.136 & 23.662 $\pm$ 0.061 & 0.81 $\pm$ 0.11          & -3.04 $\pm$ 0.38 \\
T5 & 45          & 00:20:17.43 & 59:18:39.2 &        &   & WNE         & 21.054 $\pm$ 0.044 & 20.918 $\pm$ 0.054 &                    & 19.532 $\pm$ 0.069 &                    & 1.27 $\pm$ 0.07          & -8.41 $\pm$ 0.25 \\
R9 & 60          & 00:20:20.31 & 59:18:39.5 & WNE    & c & WNE         & 21.854 $\pm$ 0.020 & 22.176 $\pm$ 0.011 & 20.572 $\pm$ 0.022 & 21.225 $\pm$ 0.022 & 21.412 $\pm$ 0.024 & 0.80 $\pm$ 0.02          & -4.94 $\pm$ 0.09 \\
T6 & 60          & 00:20:20.34 & 59:18:37.3 &        &   & WNE         & 19.939 $\pm$ 0.019 & 19.801 $\pm$ 0.027 & 18.125 $\pm$ 0.022 & 18.561 $\pm$ 0.018 & 19.222 $\pm$ 0.018 & 1.15 $\pm$ 0.03          & -9.15 $\pm$ 0.11 \\
R8 & 60          & 00:20:20.55 & 59:18:37.1 & WN10   & c & WN10        & 20.903 $\pm$ 0.010 & 20.839 $\pm$ 0.015 & 18.308 $\pm$ 0.012 & 19.714 $\pm$ 0.020 & 20.191 $\pm$ 0.009 & 0.80 $\pm$ 0.02          & -6.37 $\pm$ 0.09 \\
M7 & 66          & 00:20:21.95 & 59:17:41.0 & WC4-5  & c & WC4-5       & 20.033 $\pm$ 0.017 &                    &                    & 20.887 $\pm$ 0.049 &                    & \textit{0.92 $\pm$ 0.26} & -6.80 $\pm$ 0.87 \\
M9 &             & 00:20:22.67 & 59:18:46.6 & WN3    & c & WN3         & 22.284 $\pm$ 0.007 & 23.758 $\pm$ 0.013 & 21.622 $\pm$ 0.026 & 22.822 $\pm$ 0.065 & 22.969 $\pm$ 0.030 & 0.79 $\pm$ 0.05          & -3.24 $\pm$ 0.19 \\
R11& (71)        & 00:20:22.74 & 59:17:53.2 & WC4    & c & WC4         & 22.615 $\pm$ 0.012 & 24.318 $\pm$ 0.030 &                    &                    & 23.385 $\pm$ 0.064 & \textit{0.92 $\pm$ 0.26} & -2.71 $\pm$ 0.87 \\
M10&             & 00:20:23.31 & 59:17:42.0 & WC7    & c & WC7         & 19.548 $\pm$ 0.010 & 21.602 $\pm$ 0.011 & 19.261 $\pm$ 0.013 & 20.012 $\pm$ 0.017 & 20.481 $\pm$ 0.007 & 1.26 $\pm$ 0.02          & -7.03 $\pm$ 0.09 \\
R12& 97          & 00:20:25.65 & 59:16:48.1 & WNE    & c & WNE         & 22.677 $\pm$ 0.016 & 23.552 $\pm$ 0.017 & 20.251 $\pm$ 0.041 & 21.920 $\pm$ 0.041 & 22.587 $\pm$ 0.025 & 1.34 $\pm$ 0.04          & -5.44 $\pm$ 0.14 \\
M12& 100         & 00:20:26.19 & 59:17:26.2 & WC4    & c & WC4         & 21.466 $\pm$ 0.014 & 22.310 $\pm$ 0.010 & 19.858 $\pm$ 0.052 & 20.302 $\pm$ 0.039 & 21.256 $\pm$ 0.028 & 1.63 $\pm$ 0.03          & -8.09 $\pm$ 0.13 \\
R10& 106         & 00:20:26.51 & 59:17:04.9 & WC4    & c & WC4         & 22.486 $\pm$ 0.014 & 24.462 $\pm$ 0.027 & 19.115 $\pm$ 0.081 &                    & 22.726 $\pm$ 0.049 & \textit{0.92 $\pm$ 0.26} & -2.31 $\pm$ 0.87 \\
M13& 111         & 00:20:26.66 & 59:17:32.6 & WC5-6  & c & WC5-6       & 20.898 $\pm$ 0.012 & 21.259 $\pm$ 0.009 & 19.903 $\pm$ 0.036 & 20.577 $\pm$ 0.020 & 20.713 $\pm$ 0.014 & 0.59 $\pm$ 0.02          & -5.22 $\pm$ 0.09 \\
M14& (111)       & 00:20:26.90 & 59:17:19.7 & WC5    & c & WC5         & 20.165 $\pm$ 0.007 & 21.287 $\pm$ 0.006 & 20.266 $\pm$ 0.059 & 20.117 $\pm$ 0.025 & 20.569 $\pm$ 0.008 & 0.97 $\pm$ 0.02          & -6.59 $\pm$ 0.09 \\
M15&             & 00:20:27.06 & 59:18:17.4 & WC6-7  & a & WC4         & 21.726 $\pm$ 0.006 & 23.529 $\pm$ 0.031 & 21.672 $\pm$ 0.044 & 22.923 $\pm$ 0.079 & 22.477 $\pm$ 0.021 & 0.53 $\pm$ 0.07          & -3.04 $\pm$ 0.24 \\
M24& 111c        & 00:20:27.70 & 59:17:37.1 & WN/OB  & c & O2.5 If/WN6 & 19.077 $\pm$ 0.019 & 19.052 $\pm$ 0.020 & 18.240 $\pm$ 0.039 & 18.438 $\pm$ 0.022 & 18.486 $\pm$ 0.014 & 0.65 $\pm$ 0.02          & -8.36 $\pm$ 0.10 \\
R2 & (106/115)   & 00:20:28.03 & 59:17:14.1 & WN7-8  & c & WN7-8       & 21.770 $\pm$ 0.013 & 22.083 $\pm$ 0.012 & 19.834 $\pm$ 0.013 & 20.867 $\pm$ 0.013 & 21.477 $\pm$ 0.016 & 0.97 $\pm$ 0.01          & -5.98 $\pm$ 0.08 \\
M17&             & 00:20:29.09 & 59:16:51.7 & WNE +BH& c & WNE +BH     & 22.747 $\pm$ 0.012 & 23.194 $\pm$ 0.011 & 22.164 $\pm$ 0.098 & 22.119 $\pm$ 0.034 & 22.590 $\pm$ 0.031 & 0.90 $\pm$ 0.03          & -5.66 $\pm$ 0.12 \\
M19&             & 00:20:31.04 & 59:19:04.5 & WN4    & c & WN4         & 22.095 $\pm$ 0.007 & 23.220 $\pm$ 0.011 & 21.288 $\pm$ 0.025 & 22.616 $\pm$ 0.055 & 22.467 $\pm$ 0.018 & 0.53 $\pm$ 0.05          & -3.28 $\pm$ 0.17 \\
M23& 139         & 00:20:32.76 & 59:17:16.2 & WN7-8  & b & WN7         & 22.195 $\pm$ 0.009 & 22.521 $\pm$ 0.014 & 19.432 $\pm$ 0.022 & 20.164 $\pm$ 0.010 & 21.660 $\pm$ 0.013 & \textit{0.92 $\pm$ 0.26} & -5.41 $\pm$ 0.87 \\
M20&             & 00:20:34.49 & 59:17:14.4 & WC5    & c & WC5         & 21.313 $\pm$ 0.005 & 23.246 $\pm$ 0.014 & 21.153 $\pm$ 0.031 & 22.186 $\pm$ 0.042 & 22.223 $\pm$ 0.013 & 0.88 $\pm$ 0.04          & -4.62 $\pm$ 0.14 \\
M21&             & 00:20:41.62 & 59:16:24.4 & WN4    & c & WN4         & 22.031 $\pm$ 0.007 & 23.445 $\pm$ 0.017 & 21.262 $\pm$ 0.021 & 22.424 $\pm$ 0.033 & 22.666 $\pm$ 0.023 & 0.86 $\pm$ 0.03          & -4.21 $\pm$ 0.12 \\
\hline
\hline
\multicolumn{14}{l}{
  \begin{minipage}{\textwidth}~\\
    Parentheses around HL regions indicate the WR star is in the close vicinity of, but not necessarily within, the H\,{\sc ii} region in question. \\
    Italics denote a WR star with an E(B-V) value derived from an average, rather than individually calculated (see Sect.~\ref{sec:ebv}). \\
    Note the minor revisions to the spectral type of M15 - updated from WC6-7 \citep{mas1995}, M23 - updated from WN7-8, and M24 - updated from WN/OB \citep{mas2002a, cro2003}. \\
    References as follows: a: \citet{mas1995}; b: \citet{mas2002a}; c: \citet{cro2003}; \\
  \end{minipage}
}\\
\end{tabular}
\end{center}
\end{table}
\end{landscape}


\subsection{Spectroscopy}
\label{sec:spectroscopy} 

We obtained follow-up spectroscopic observations, which took place on 10-11 September 2010, again using GMOS on Gemini-North with the program ID GN-2010B-Q-44 (PI Crowther). Two masks were constructed, (mask 1 and 2), each containing 20 targets, with their properties detailed in Table~\ref{tab:masks}. Of the 9 potential candidates identified in Table~\ref{tab:potentialWR}, four were included (T4,T5,T6 and T9) across the two masks.

\begin{table}
\caption{Properties of the IC10 masks used with GMOS instrument for multi-object spectroscopy, where mask 1 and 2 correspond to the most recent data set and mask 3 and 4 are from \citet{cro2003}. For all cases the slit length was set at 5$\arcsec$.}
\label{tab:masks}
\begin{tabular}{c@{\hspace{3mm}}c@{\hspace{3mm}}c@{\hspace{3mm}}c@{\hspace{3mm}}c@{\hspace{3mm}}}
\hline
\hline
Mask & Slit Width & Observation & FWHM \\
 & [$\arcsec$] & Date & [$\arcsec$] \\
\hline
1 & 0.75 & 10 Sept 2010              & 0.5 \\
2 & 0.75 & 10 Sept 2010 - 11 Sept 2010 & 0.6 \\
3 & 0.8  & 22 Dec 2001 - 16 Jan 2002 & 0.7 \\
4 & 0.8  & 22 Dec 2001 - 15 Jan 2002 & 0.8 \\
\hline
\hline
\end{tabular}
\end{table}

Four exposures per mask were obtained, each for a duration of 2600s. A slit width was fixed at 0.75$\arcsec$ and the B600 grating was used in all cases, and the spectral resolution of the data was found to be 3.4 \ang\; from arc lines. To compensate for the gaps present in the GMOS detector, the central wavelength was shifted by 20nm from 510nm to 530nm for one pair of exposures for each mask. Signal to noise was improved by merging the four exposures per target, and targets common to both masks were also combined.

Wavelength calibration was completed using an internal CuAr arc lamp, and for the flux calibration, the white dwarf star G191B2B was observed through the same B600 grating with a slit width of 0.75$\arcsec$. Total integration time was 3 $\times$ 30s with a shift in central wavelength between exposures from 410nm, 510nm, and 610nm to ensure the generated response function covered the necessary wavelength range. Slit loss corrections were also applied from photometric magnitudes.

This data set was combined with previous GMOS observations, GN-2001B-Q-20 (mask 3) and GN-2001B-Q-23 (mask 4), which were obtained between 21 December 2001 and 16 January 2002 \citep{cro2003} (see Table~\ref{tab:masks}). Those WR observed in multiple masks underwent further merging to produce a single spectra for each candidate. 


\section{Nebular Results}
\label{sec:neb_res}


\subsection{Nebular Extinction}
\label{sec:neb_ext}

A number of spectra within the datasets exhibited prominent nebular emission, prompting a separate analysis to investigate the gaseous properties of IC10. Known H\,{\sc ii} regions were identified using the maps produced by \citet{hod1990}, and we also include data from candidate H\,{\sc ii} regions suggested by P. Royer (private communication). Table~\ref{tab:nebularflux} shows which regions were included and provides an overview of the available emission line fluxes measured for each region, relative to \hbeta = 100.

Individual reddening corrections, based on Balmer emission line ratios, were computed for each H\,{\sc ii} region. Depending on the available wavelength range, \halpha/\hbeta or \hgamma/\hbeta ratios were used in conjunction with an intrinsic intensity ratio to obtain a measure of c(\hbeta). Observed emission flux measurements were corrected for underlying stellar absorption using:
\begin{equation*}
  f_{\lambda corr}=f_{0}\frac{W_{\lambda} + W_{abs}}{W_{\lambda}} 
  \label{eq:stellarabsorptioncor}
\end{equation*}
where $f_{0}$ refers to the observed flux, $f_{\lambda corr}$ is the corrected flux and values for W$_{abs}$ were taken from \citet{gon1999}, for an instantaneous burst with Salpeter IMF, mass range of 1-80 \msol\ and age of 2 Myrs. By interpolating between varying metallicity intervals, the W$_{abs}$ parameter was determined to be 2.5\ang, 2.4\ang and 2.5\ang for \halpha, \hbeta and \hgamma respectively. Measured equivalent widths (W$_{\lambda}$) for all Balmer emission lines are included in Table~\ref{tab:nebularew} in the Appendix.

The corrected fluxes were then used with:
\begin{equation*}
    \frac{f_{\lambda corr}}{f_{\beta corr}}=\frac{I_{\lambda}}{I_{\beta}}10^{c(H\beta)[X_{\lambda}- X_{\beta}]} 
    \label{eq:Hextinction}
\end{equation*}
where X$_{H\alpha}$, X$_{H\beta}$ and X$_{H\gamma}$ are 0.82, 1.17 and 1.32 respectively, determined from a Galactic extinction law \citep{sea1979}, and the intrinsic intensity ratios are $\frac{I(H\alpha)}{I(H\beta)}$=2.86 and $\frac{I(H\gamma)}{I(H\beta)}$=0.47. Table~\ref{tab:nebularflux} shows the derived c(\hbeta) measurements for each H\,{\sc ii} region, which when combined give an average nebular extinction of c(\hbeta)=1.19 $\pm$ 0.28, or E(B-V) $\sim$ 0.7c(\hbeta) = 0.83 $\pm$ 0.20. \citet{sch2011a} find the Milky Way foreground contribution in the direction of IC10 to be E(B-V)=1.39, however note the low galactic latitude of the galaxy translated to a highly uncertain extinction estimate.

\begin{table*}
\begin{center}
  \caption{Nebular emission line flux measurements relative to H$\beta$=100 for various H\,{\sc ii} regions. HL\# refer to H\,{\sc ii} regions outlined by \citet{hod1990} and H\,{\sc ii}\# refer to candidate H\,{\sc ii} regions suggested by \citet{roy2001}. Final column indicates the reddening correction derived for that region using Balmer emission line ratios as described in the text.}
  \label{tab:nebularflux} \begin{tabular}{c@{\hspace{1.5mm}}c@{\hspace{1.5mm}}c@{\hspace{1.5mm}}c@{\hspace{1.5mm}}c@{\hspace{1.5mm}}c@{\hspace{1.5mm}}c@{\hspace{1.5mm}}c@{\hspace{1.5mm}}c@{\hspace{1.5mm}}c@{\hspace{1.5mm}}c@{\hspace{1.5mm}}c@{\hspace{1.5mm}}c@{\hspace{1.5mm}}}
\hline
\hline
Nebular & Mask & \lbrack O\,{\sc ii}\rbrack & H$\gamma$ & \lbrack O\,{\sc iii}\rbrack & H$\beta$ & \lbrack O\,{\sc iii}\rbrack & \lbrack O\,{\sc iii}\rbrack & \lbrack N\,{\sc ii}\rbrack & H$\alpha$ & \lbrack N\,{\sc ii}\rbrack & $10^{-17}$ F$_{H\beta}$ & c(\hbeta) \\
Region  &  & 3727 & 4340 & 4363 & 4861 & 4959 & 5007 & 6548 & 6562 & 6584 & [\ergcms] & \\
\hline
HL 6          &2&            &              &               & 100 $\pm$ 17 & 251 $\pm$ 22  & 801 $\pm$ 25 & 64 $\pm$ 15 & 1035 $\pm$ 16 & 172 $\pm$ 15 & 1.89 $\pm$ 0.31 & 1.06\\
HL 10         &2&            &              &               & 100 $\pm$ 8  & 64 $\pm$ 8    & 217 $\pm$ 9  & 41 $\pm$ 8  & 1117 $\pm$ 11 & 117 $\pm$ 8  & 2.51 $\pm$ 0.21 & 1.53\\
HL 20         &3&            & 39 $\pm$ 4   &               & 100 $\pm$ 5  &               &              & 33 $\pm$ 6  & 960 $\pm$ 8   & 72 $\pm$ 6   & 16 $\pm$ 0.76   & 1.29\\
HL 22         &4&            &              &               & 100 $\pm$ 1  & 116 $\pm$ 3   & 350 $\pm$ 3  & 9 $\pm$ 4   & 640 $\pm$ 5   & 26 $\pm$ 4   & 109 $\pm$ 0.89  & 0.99\\
HL 45         &3&            &              &               & 100 $\pm$ 1  & 174 $\pm$ 2   & 508 $\pm$ 3  & 10 $\pm$ 3  & 628 $\pm$ 3   & 29 $\pm$ 3   & 45 $\pm$ 1.1    & 0.96\\
HL 45         &1& 21 $\pm$ 1 & 26 $\pm$ 0.2 & 2.2 $\pm$ 0.2 & 100 $\pm$ 2  & 222 $\pm$ 2   &              &             &               &              & 381 $\pm$ 5.8   & 1.62\\
H\,{\sc ii} 04&4&            & 42 $\pm$ 5   &               & 100 $\pm$ 3  & 115 $\pm$ 3   & 367 $\pm$ 4  &             & 555 $\pm$ 31  & 27 $\pm$ 23  & 6.56 $\pm$ 0.21 & 0.61\\
H\,{\sc ii} 07&3&            & 30 $\pm$ 3   &               & 100 $\pm$ 1  & 135 $\pm$ 2   & 449 $\pm$ 3  & 10 $\pm$ 3  & 731 $\pm$ 3   & 28 $\pm$ 3   & 127 $\pm$ 1.71  & 1.18\\
H\,{\sc ii} 07&4&            &              &               & 100 $\pm$ 3  & 132 $\pm$ 2   & 413 $\pm$ 3  & 11 $\pm$ 3  & 719 $\pm$ 4   & 36 $\pm$ 3   & 137 $\pm$ 3.72  & 1.16\\
H\,{\sc ii} 08&3&            &              &               & 100 $\pm$ 1  & 135 $\pm$ 1   & 428 $\pm$ 2  & 13 $\pm$ 3  & 809 $\pm$ 3   & 42 $\pm$ 3   & 148 $\pm$ 0.80  & 1.31\\
H\,{\sc ii} 08&4& 41 $\pm$ 2 & 29 $\pm$ 0.4 &               & 100 $\pm$ 1  & 141 $\pm$ 1   & 439 $\pm$ 2  &             &               &              & 213 $\pm$ 2.83  & 1.36\\
H\,{\sc ii} 11&3&            & 7 $\pm$ 1    &               & 100 $\pm$ 2  & 150 $\pm$ 6  & 500 $\pm$ 7 & 37 $\pm$ 5    & 2041 $\pm$ 7  & 113 $\pm$ 5  & 22 $\pm$ 0.36   & -   \\
\hline
\hline
\multicolumn{13}{l}{
  \begin{minipage}{0.93\textwidth}~\\
     RA and DEC (J2000) co-ordinates for \citet{roy2001} H\,{\sc ii} regions as follows: H\,{\sc ii} 04 (00:20:15.48, +59:18:40.6) H\,{\sc ii} 07 (00:20:18.51, +59:17:40.4) H\,{\sc ii} 08 (00:20:24.41, +59:16:55.2) H\,{\sc ii} 11 (00:20:19.36, +59:18:02.9) \\
  \end{minipage}
}\\
  \end{tabular}
  \end{center}
\end{table*}


\subsection{Metallicity}
\label{sec:met}

Metallicity measurements are important for providing information about the local environment, and the star formation history of the galaxy. A previous metallicity determination for two H\,{\sc ii} regions, IC10-1 and IC10-2, outlined by \citet{leq1979} and later catalogued as HL111 and HL45 respectively \citep{hod1990}, suggested oxygen abundance measurements of log(O/H) + 12 = 8.17 and log(O/H) + 12 = 8.45 respectively. Unfortunately the quality of the [O\,{\sc iii}] 4363 intensity measurement was flagged as uncertain. \citet{gar1990} performed a second analysis of HL45, obtaining a metallicity of log(O/H) + 12 = 8.26, again based on the observations from \citet{leq1979}. \citet{ric2001} also report metallicity measurements for the H\,{\sc ii} regions HL111b and HL111c located within the IC10-1 region and find log(O/H) + 12 = 7.84 and log(O/H) + 12 = 8.23 respectively. These results are summarised in Table~\ref{tab:HLneb}.

Using the nebular emission present in the regions outlined in Table~\ref{tab:nebularflux}, oxygen abundance measurements based on strong line methods using the N2 and O3N2 ratios from \citet{pet2004} were determined and are shown in Table~\ref{tab:o3n2}. Unfortunately, the linear relationship between oxygen abundance and the O3N2 ratio only holds true for O3N2 measurements within the range of -1 to 1.9, and for IC10 our measured O3N2 ratios suggest we are at the limit of this calibration beyond which this linear relationship breaks down. Nevertheless, using our O3N2 ratio we find an average metallicity of log(O/H) + 12 = 8.14 $\pm$ 0.09, with a systematic uncertainty from the method of 0.25. For completeness we also compute oxygen abundance measurements using the N2 ratio, for which we find log(O/H) + 12 = 8.22 $\pm$ 0.14, however we note that the N2 method produces results with a large dispersion, therefore introducing a systematic uncertainty on the metallicity of $\pm$ 0.41.

\begin{table}
\caption{Oxygen abundance measurements for H\,{\sc ii} regions within IC10, derived using the N2 and O3N2 strong line methods outlined in \citet{pet2004}}
\label{tab:o3n2}
\begin{tabular}{c@{\hspace{1.5mm}}c@{\hspace{1.5mm}}c@{\hspace{1.5mm}}c@{\hspace{1.5mm}}c@{\hspace{1.5mm}}c@{\hspace{1.5mm}}}
\hline
\hline
Nebular & Mask & N2 & log(O/H) + 12 & O3N2 & log(O/H) + 12 \\
Region  &      &    &               &      &               \\
\hline
HL 6   & 2 & -0.78 & 8.46 $\pm$ 0.41 & 1.68 & 8.19 $\pm$ 0.26 \\
HL 10  & 2 & -0.98 & 8.34 $\pm$ 0.41 & 1.32 & 8.31 $\pm$ 0.26 \\
HL 22  & 4 & -1.39 & 8.11 $\pm$ 0.42 & 1.93 & 8.11 $\pm$ 0.26 \\
HL 45  & 3 & -1.34 & 8.14 $\pm$ 0.41 & 2.04 & 8.08 $\pm$ 0.25 \\
H\,{\sc ii} 04 & 4 & -1.31 & 8.15 $\pm$ 0.55 & 1.87 & 8.13 $\pm$ 0.45 \\
H\,{\sc ii} 07 & 3 & -1.42 & 8.09 $\pm$ 0.41 & 2.07 & 8.07 $\pm$ 0.25 \\
H\,{\sc ii} 07 & 4 & -1.31 & 8.16 $\pm$ 0.41 & 1.92 & 8.11 $\pm$ 0.25 \\
H\,{\sc ii} 08 & 3 & -1.29 & 8.17 $\pm$ 0.41 & 1.92 & 8.12 $\pm$ 0.25 \\
H\,{\sc ii} 11 & 3 & -1.26 & 8.18 $\pm$ 0.41 & 1.96 & 8.10 $\pm$ 0.25 \\
\hline
\hline
\end{tabular}
\end{table}

\begin{figure*}
	\includegraphics[width=\textwidth]{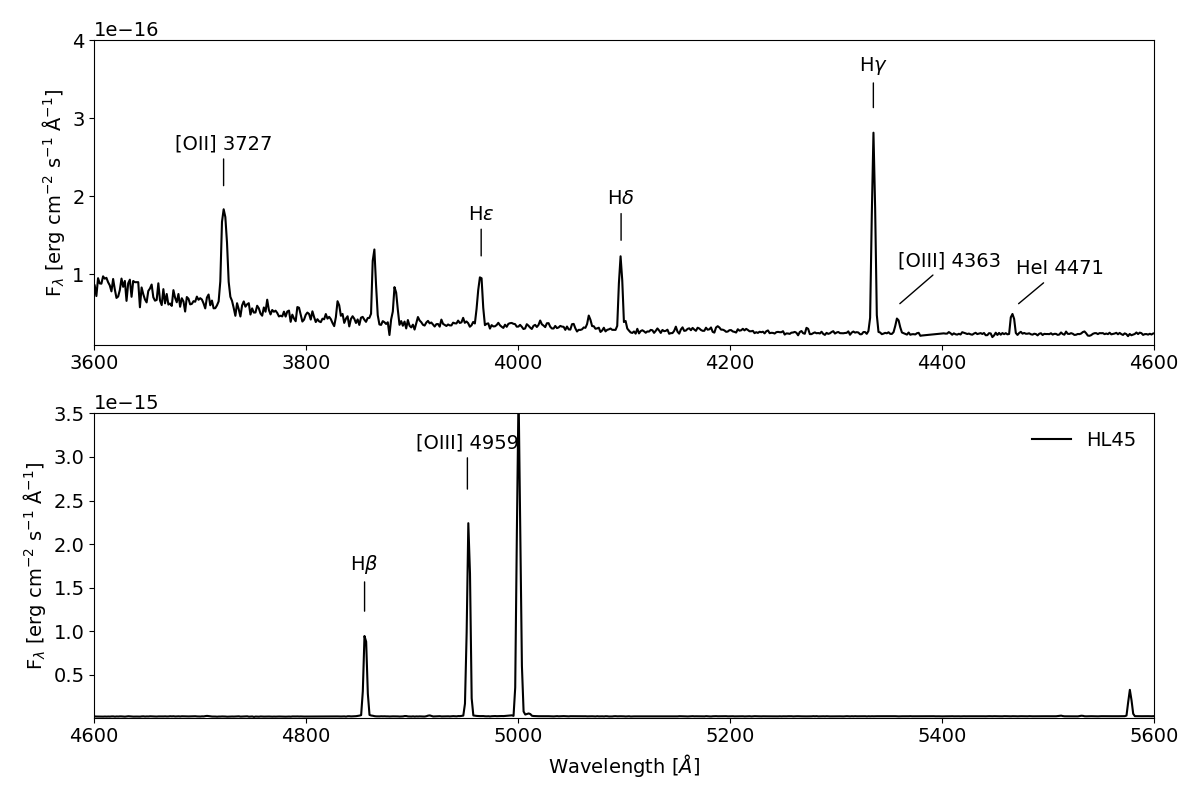}
    \caption{Flux calibrated optical nebular spectra of the H\,{\sc ii} region HL45, associated with the WR star T5. Labels highlight the Balmer series and other forbidden lines important for metallicity measurements, including a clear [O\,{\sc iii}] 4363 emission line.}
    \label{fig:nebt5}
\end{figure*}

To produce a more robust determination of the oxygen content we also derive an updated metallicity measurement for IC10 using the nebular emission spectra of the newly confirmed WR star T5, which is associated with the HL45 H\,{\sc ii} region \citep{hod1990} and is shown in Fig.~\ref{fig:nebt5}. Whilst nebular emission was present in a number of spectra, solely T5 provided a robust [O\,{\sc iii}] 4363 flux measurement, as shown in Table~\ref{tab:nebularflux}, necessary when calculating metallicity using the direct $T_{e}$ method.

HL45 was corrected for an extinction of c(\hbeta)=1.62 $\pm$ 0.05 found using the method outlined in Sect.~\ref{sec:neb_ext} before appropriate emission line intensities were measured. An [O\,{\sc iii}] electron temperature of 9700$\pm$250 K was found using the nebular package in IRAF, and the same temperature was adopted for the [O\,{\sc ii}] gas temperature. This approximation should be sufficient since the contribution from the [O\,{\sc iii}] region dominates the final oxygen abundance measurement, as shown in Table~\ref{tab:HLneb}. A change in $\pm$ 1000K in [O\,{\sc ii}] gas temperature corresponds to a $\pm$ 0.03 adjustment in log(O/H) + 12.

Assuming a density of 100 cm$^{-3}$ gave an oxygen abundance of log(O/H) + 12 = 8.40 $\pm$ 0.04. Present results are included in Table~\ref{tab:HLneb}. It is apparent that our updated oxygen abundance for HL 45 is similar to \citet{leq1979} although is somewhat higher than both \citet{gar1990} and \citet{ric2001}. We consider this oxygen measurement to be a good representation of the global metallicity of IC10 because the oxygen content distribution for other dwarf galaxies has been shown to be relatively uniform. Integral field studies of blue compact dwarf (BCD) galaxies \citep{gar2008a, cai2015a} find that for a BCD with a metallicity greater than 8.1, the variation in log(O/H) + 12 across the galaxy does not exceed $\sim$0.1 dex. A slightly higher metallicity for IC10 is still consistent with the luminosity-metallicity relationship from \citet{shi2005a} for M$_{B}$ = $-$16.3 mag\footnote{Absolute blue magnitude for IC10 obtained from m$_{B}$ = 11.8 mag, A$_{B}$ = 4.1 $\times$ E(B-V) = 3.8 mag and a distance modulus of 24.3 mag.}. The oxygen content of this galaxy is therefore more similar to the LMC (log(O/H) + 12 = 8.37) than the SMC (log(O/H) + 12 = 8.13), and IC10 is not as metal-poor as previously considered.

\begin{table}
\caption{Overview of previously derived [O\,{\sc iii}] temperatures and metallicity measurements for two different H\,{\sc ii} regions within IC10 (IC10 1 and IC10 2), outlined by \citet{leq1979}, and present results from mask1 observations of HL45}
\label{tab:HLneb}
\begin{tabular}{c@{\hspace{0mm}}c@{\hspace{0.5mm}}c@{\hspace{0.5mm}}c@{\hspace{0.5mm}}c@{\hspace{0.5mm}}c@{\hspace{-3mm}}c@{\hspace{0mm}}}
\hline
\hline
Lequeux & HL     & T(O$^{2+}$)         & (O$^{+}$/H)       & (O$^{2+}$/H)        & log(O/H) & Ref\\
Region  &        & [$\times 10^{4}$ K] & [$\times 10^{5}$] & [$\times 10^{4}$]  & + 12     &     \\
\hline
IC10 1  & 111    & 1.16             & 3.98            & 1.10            & 8.17            & a \\
IC10 1  & 111b   & 1.40 $\pm$ 0.30  &                 &                 & 7.84 $\pm$ 0.25 & b \\
IC10 1  & 111c   & 1.00 $\pm$ 0.06  &                 &                 & 8.23 $\pm$ 0.09 & b \\
IC10 2  & 45     & 1.06             & 6.03            & 2.19            & 8.45            & a \\
IC10 2  & 45     & 1.08             &                 &                 & 8.26 $\pm$ 0.10 & c \\
IC10 2  & 45     & 0.97 $\pm$ 0.03  & 2.28 $\pm$ 0.19 & 2.30 $\pm$ 0.21 & 8.40 $\pm$ 0.04 & d \\
\hline
\hline
\multicolumn{7}{l}{
  \begin{minipage}{\columnwidth}~\\
    a: \citet{leq1979}; b: \citet{ric2001}; c: \citet{gar1990}; d: This work; \\
  \end{minipage}
}\\
\end{tabular}
\end{table}


\subsection{Star Formation Rate}
\label{sec:sfr}

\begin{figure*}
	\includegraphics[width=2\columnwidth]{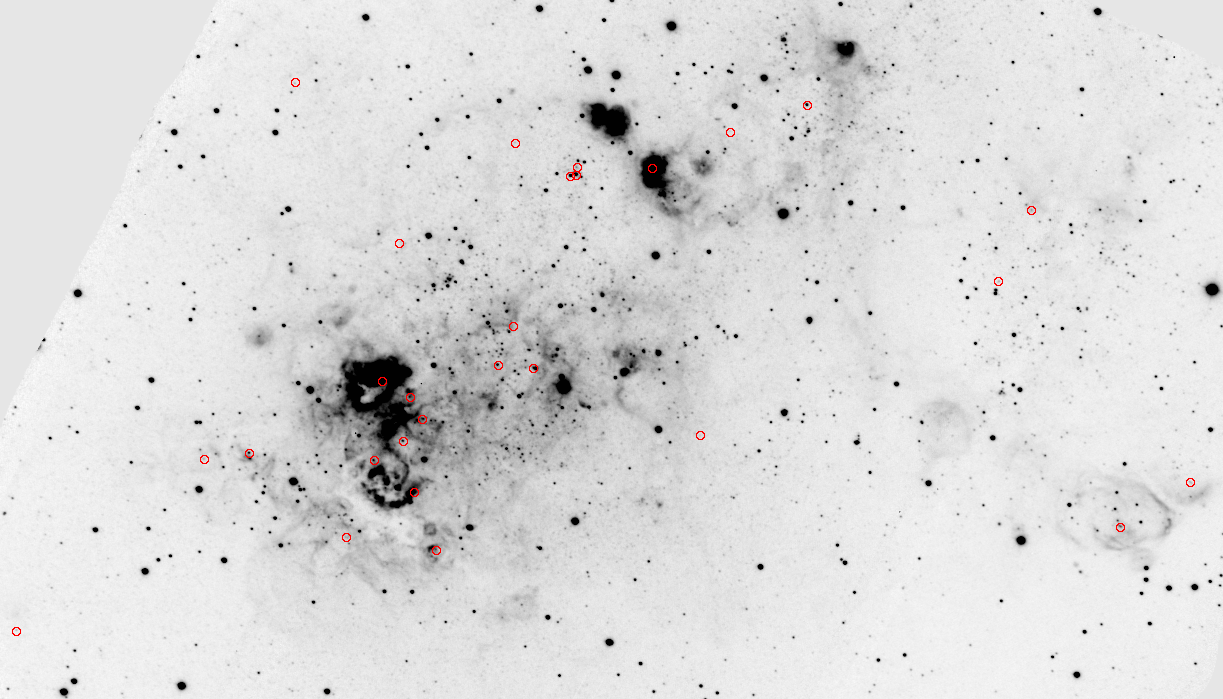}
    \caption{Gemini GMOS \halpha\ image of IC10, north is up and east is left, showing the distribution of ionized gas and the relative positions of confirmed WR stars (red circles) with the exception of M5 which is out of the field of view but is associated with the H\,{\sc ii} region HL29. The bright star forming complex HL106/111 (left of centre) contains 6 WR stars.  Field of view is 384\arcsec\ $\times$ 210\arcsec, corresponding to 1.4 $\times$ 0.8 kpc at a distance of 740 kpc.}
    \label{fig:ha}
\end{figure*}

Using the \halpha\ and \halpha C imaging discussed in Sect.~\ref{sec:photometry} we also re-determine the star formation rate (SFR) of IC10. Fig.~\ref{fig:ha} shows the \halpha\ image of IC10, and we include the positions of known WR stars for reference. Using aperture photometry to obtain a total \halpha\ count rate, we apply a conversion derived from standard star measurements to find the integrated \halpha\ flux. A similar exercise for the \halpha C filter permits the stellar contribution to be subtracted. This approach was preferred to producing a net \halpha\ - \halpha C image, which introduced subtraction artefacts due to bright stars within the field of view. To correct for [N\,{\sc ii}] emission included within the GMOS \halpha\ filter bandwidth, we apply the correction:
\begin{equation*}
  \frac{\text{F(\lbrack N\,{\sc ii}\rbrack 6548+6584)}}{\text{F(\halpha)}}=0.09 \pm 0.02
  \label{eq:nIIha}
\end{equation*}
derived from average flux measurements of the [N\,{\sc ii}] 6548, [N\,{\sc ii}] 6584 and \halpha\ emission lines across all available H\,{\sc ii} regions (shown in Table~\ref{tab:sfr}).

The average gas extinction of c(\hbeta)=1.19 $\pm$ 0.28 was applied and adopting a distance of 740 $\pm$ 20 kpc we find L$_{H\alpha}$=5.64 $\pm$ 2.93$\times$10$^{39}$ \ergs. To convert this to a SFR we use:
\begin{equation*}
  \text{SFR} = 7.94 \times 10^{-42} L_{H\alpha}
  \label{eq:sfr}
\end{equation*}
which assumes a Salpeter function IMF over a mass range of 0.1-100 \msol, and resulted in a SFR=0.045 $\pm$ 0.023 \sfr\ \citep{ken1998}. We also derive a SFR for the dominant giant H\,{\sc ii} region of IC10 comprising the complex centred on HL111/106 \citep{hod1990}. An elliptical aperture with semi-major/minor axes of 17\arcsec $\times$ 27\arcsec reveals a \halpha\ luminosity of 1.4$\times$10$^{39}$ \ergs, typical of the brightest H\,{\sc ii} regions of local star forming galaxies \citep{ken1988}. These results are shown in Table~\ref{tab:localgroup}.

A summary of the current and previous SFR determinations, using both \halpha\ and radio flux measurements, are shown in Table~\ref{tab:sfr}. \citet{ken2008} find a lower SFR using a similar method, owing to a lower \halpha\ flux and smaller dust extinction correction. In contrast, our SFR lies intermediate between the radio derived SFRs by \citet{gre1996} and \citet{chy2016}. The comparison to radio derived SFRs is useful because at these wavelengths dust extinction is negligible, and therefore does not effect the measured radio flux. We do, however, note that contributions from non-thermal radio sources such as synchrotron emission from supernova remnants can skew results, especially at longer radio wavelengths. The radio SFRs shown from \citet{gre1996} and \citet{chy2016} have not been corrected for this contribution.

\begin{table*}
  \begin{center}
\caption{Comparison of the IC10 star formation rates derived using both radio flux measurements and \halpha\ luminosity measurements, including a new attempt from this work. All values have been scaled to a distance of 740 kpc. Radio SFRs assume all flux measured is free-free radio emission and there is no contribution from non-thermal sources.}
\label{tab:sfr}
\begin{tabular}{c@{\hspace{1.5mm}}c@{\hspace{1.5mm}}c@{\hspace{1.5mm}}c@{\hspace{1.5mm}}c@{\hspace{1.5mm}}c@{\hspace{1.5mm}}c@{\hspace{1.5mm}}c@{\hspace{1.5mm}}c@{\hspace{1.5mm}}}
\hline
\hline
Method   & $\nu$ & F$_{\lambda}$      & $10^{-12}$ F$_{H\alpha}$ & A$_{H\alpha}$     & N\,{\sc ii}/\halpha & $10^{39}$ L$_{H\alpha}$ & SFR               & Ref \\
         & [GHz] & [mJy]            & [\ergcms]             & [mag]           &                     & [\ergs]              & [\sfr]            &     \\
\hline
Radio    & 4.85  & 137 $\pm$ 12     &                       &                 &                     &                      &  0.030 $\pm$ 0.003 & a \\
Radio    & 1.43  & 377 $\pm$ 11     &                       &                 &                     &                      &  0.073 $\pm$ 0.005 & b \\
\halpha  &       &                  & 103 $\pm$ 28          & 1.90            & 0.080 $\pm$ 0.008   & 3.89 $\pm$ 0.86      &  0.031 $\pm$ 0.007 & c \\
\halpha  &       &                  & 130 $\pm$ 33          & 2.06 $\pm$ 0.49 & 0.092 $\pm$ 0.023   & 5.64 $\pm$ 2.93      &  0.045 $\pm$ 0.023 & d \\
\hline
\hline
\multicolumn{8}{l}{
  \begin{minipage}{0.62\linewidth}~\\
a: \citet{gre1996}; b: \citet{chy2016}; c: \citet{ken2008}; d: This work \\
  \end{minipage}
}\\
\end{tabular}
\end{center}
\end{table*}


\section{Stellar Results}
\label{sec:res}


\subsection{New Wolf-Rayet Stars}
\label{sec:confirmedWR}

\begin{figure*}
	\includegraphics[width=\textwidth]{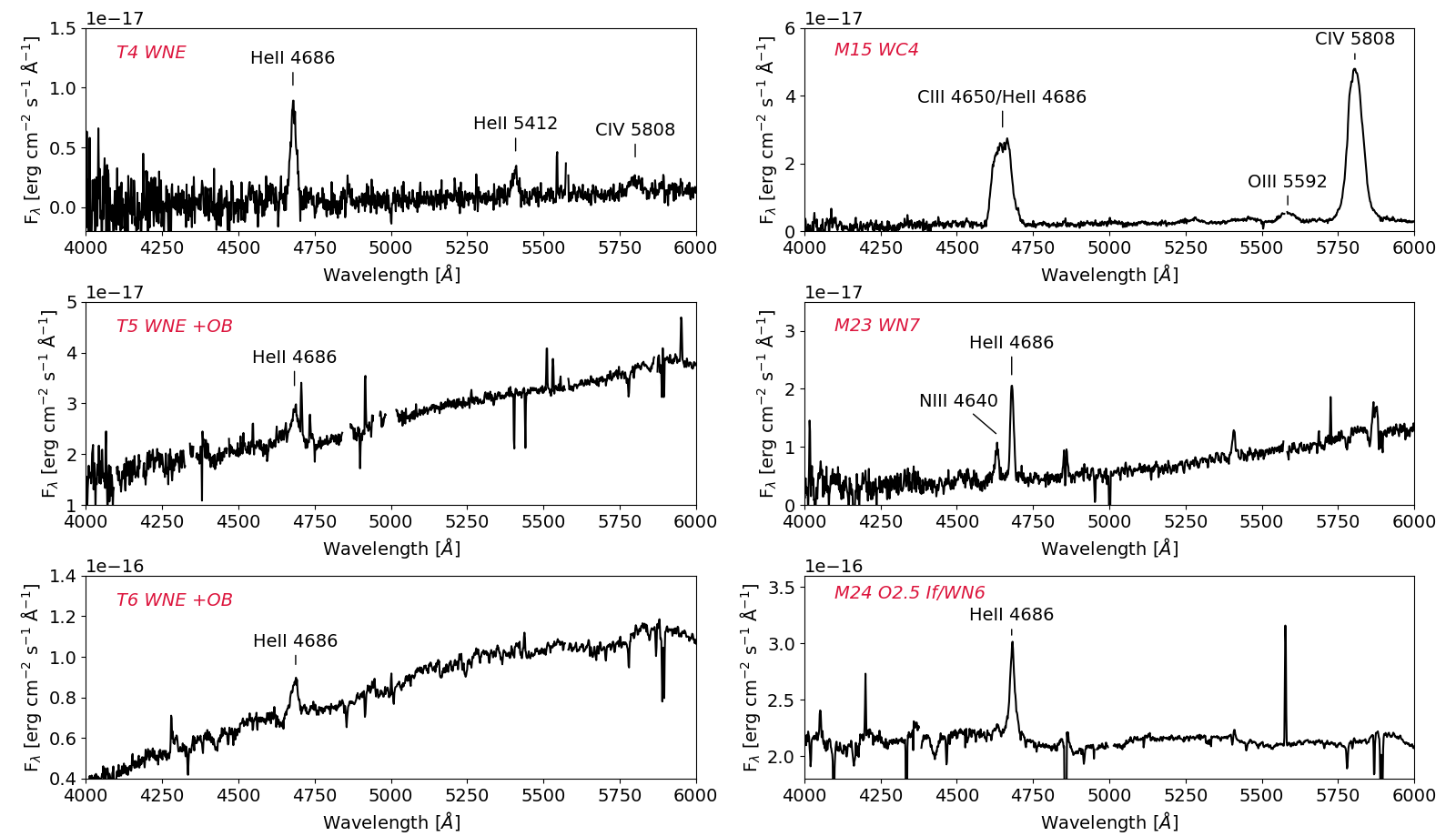}
    \caption{On the left we show the flux calibrated optical spectra of 3 newly confirmed WR stars within IC10, all belonging to the WN subclass. On the right we show the optical spectra for 3 previously confirmed WR stars, which were not included in the Gemini/GMOS datasets presented by \citet{cro2003} and therefore providing a complete set of spectra for all IC10 confirmed WR stars. Some nebular emission has been removed for clarity. The strong, broad absorption feature at $\lambda$4430 in T6 and M24 is a prominent Diffuse Interstellar Band \citep{her1995}.}
    \label{fig:newWRspectra}
\end{figure*}

The spectra of three newly confirmed WR stars in IC10 are presented in Fig.~\ref{fig:newWRspectra}. All three have been assigned early WN (WNE) spectral types. Both T5 and T6 are associated with star clusters which is evident from both the photometry and the presence of a strong continuum. Also, as discussed in Sect.~\ref{sec:met}, T5 is located within the H\,{\sc ii} region HL45. Spectroscopy for T9 was also performed and analysis concluded that it was not a WR star. The presence of molecular TiO bands suggested it is most likely to be a foreground early M-dwarf star instead.

The number of spectroscopically confirmed WR stars in IC10 has increased from 26 to 29 stars, and a summary of their spectral properties can be found in Table~\ref{tab:el}. Of the previously confirmed WR stars, we suggest some changes to prior spectral classifications, including the revision of M15 from WC6-7 \citep{mas1995} to WC4 due to the absence of C\,{\sc iii} 5696 and presence of O\,{\sc iii} 5592 shifting the equivalent width ratios outlined in \citet{smi1990, cro1998} into the WC4 category. The adjustment of M24 from a WNE/OB \citep{mas2002a} to a O2.5 If/WN6 due to its similarity to HD93162 \citep{cro2011}. Finally, the minor adjustment to the subclass of M23 from WN7-8 \citep{mas2002a} to WN7 based on the strengths of He\,{\sc i} 5876 to He\,{\sc ii} 5411. Table~\ref{tab:photdata} includes a complete census of the WR population and their classifications.

As mentioned previously, the WC/WN ratio in IC10 is peculiar, and since the addition of three new WN stars reduces the ratio from 1.3 to 1, IC10 is still regarded as an anomaly given the WC/WN ratio of $\sim$0.2 for the LMC and 0.1 for the SMC.


\subsection{Binary Fraction}
\label{sec:bf} 

Recent evidence suggests the local OB binary frequency is high. Analysis of radial velocity variations in O-stars within the 30 Doradus region of the LMC by \citet{san2013} revealed a lower limit intrinsic binary fraction of 51 $\pm$ 4\%. Similarly, using O stars residing in Galactic clusters, \citet{san2012b} found a lower limit binary fraction of 69 $\pm$ 9\%. Initial binary fractions for these regions most likely will have been higher but will have been disrupted over time.

If we assume a similar scenario for IC10, we would expect that some of these OB binaries would have survived the transition to WR stars, and therefore we expect a relatively high WR binary frequency. To assess this we outline a number of criteria used to identify potential binary candidates and compare the resultant binary fraction with that of WR stars in other Local Group galaxies.

The simplest method involves identifying absorption features from OB companions in the spectrum of each WR star. There is however, some ambiguity in this approach, since WR stars may contain intrinsic absorption features, so alone this is insufficient to confidently suggest a binary system.

Another method involves searching for unusual radial velocities in the strongest emission line features present in the spectrum. Geocentric radial velocity measurements were found for those WR stars with nebular emission available, and heliocentric corrections were found for each mask using the BCVcor task in IRAF. For those WR without nebular emission an average heliocentric radial velocity of -347 $\pm$ 75 kms$^{-1}$ was assumed. This agrees well with the expected radial velocity of 348 $\pm$ 1 kms$^{-1}$ \citep{huc1999a} and takes into consideration the average dispersion velocity of the gas, found to be 34 $\pm$ 5 kms$^{-1}$ \citep{mcc2012}. By measuring the radial velocity for He\,{\sc ii} 4686 (WN) and C\,{\sc iv} 5808 (WC) emission lines in each spectrum and comparing with the expected radial velocity measurement, the WR stars with shift excesses greater than 2$\sigma$ were identified as potential binary candidates or runaway stars.

\begin{figure*}
	\includegraphics[width=\textwidth]{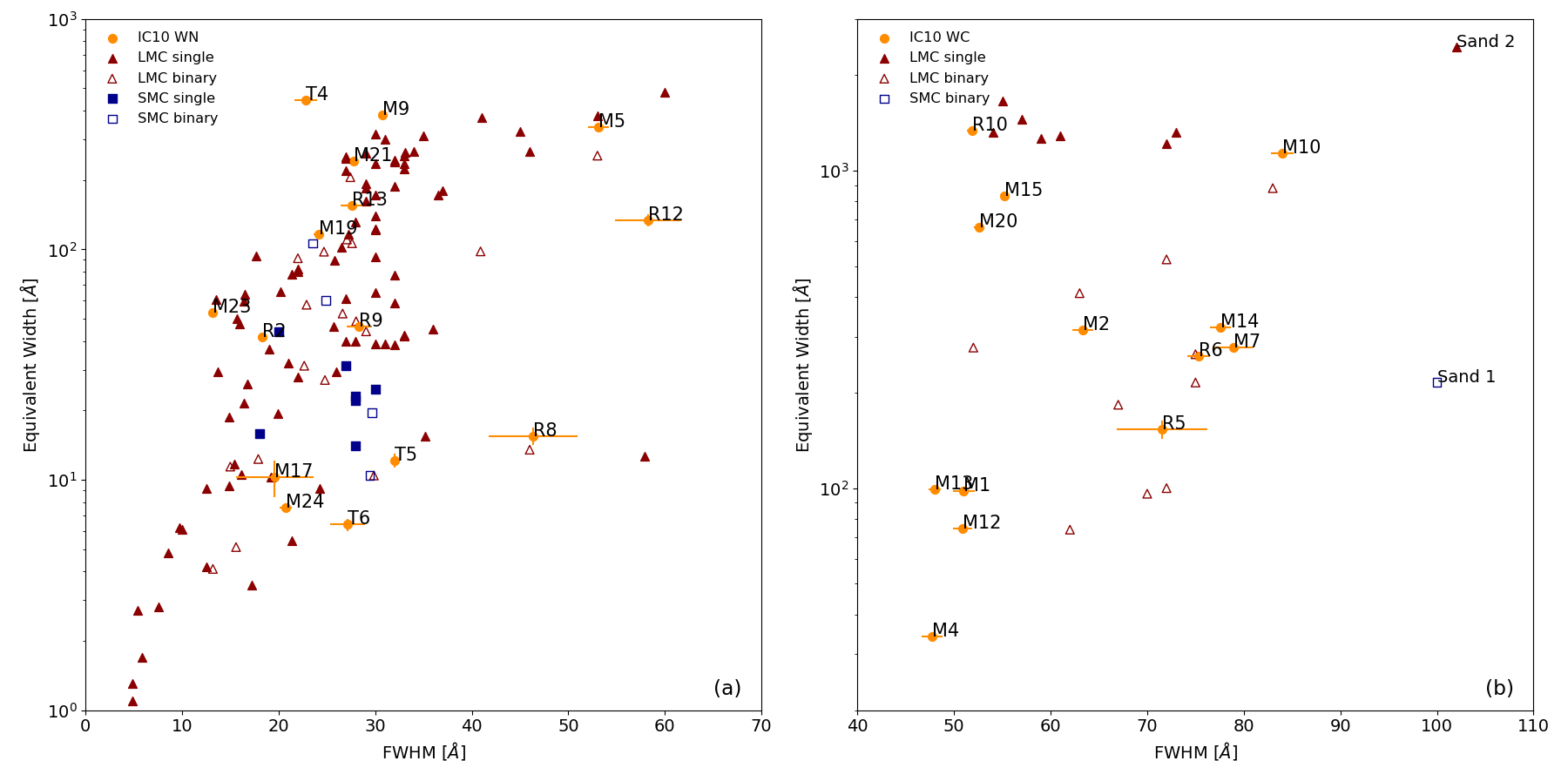}
        \caption{Shows the relationship between FWHM and equivalent width for (a) WN stars and (b) WC stars in the LMC (red triangles), SMC (blue squares), and IC10 (yellow circles). For the Magellanic cloud WR stars, binary and single stars can also be distinguished between by open and filled plot symbols respectively. The two known Magellanic WO stars have also been included (Sand 1,2). LMC data taken from \citet{cro2006}, \citet{sch2008}, and \citet{foe2003b}. SMC data from \citet{foe2003a} and \citet{cro2006}.}
    \label{fig:FWHMEW}
\end{figure*}

Finally, the continuum of an O-star companion can also dilute the emission line strength of the WR, therefore the presence of a companion would result in weak equivalent width measurements and small $\Delta$He\,{\sc ii}-He\,{\sc ii}C excesses. These stars are easily identifiable in Fig.~\ref{fig:FWHMEW} (a) and (b), in which emission line widths are compared against line strengths in IC10 and Magellanic Clouds WN and WC stars respectively. It is however, important to note that nearby sources or line-of-sight contaminations can also enhance the continuum and falsely suggest a binary system. 

Taking all these indicators into consideration, those WR stars which meet a minimum of two out of the three requirements were deemed likely to be part of a binary system, and have been indicated in Table~\ref{tab:el}, giving a coarse binary fraction of 41\%. Of these potential binary systems, four WR stars (M1, M4, T5 and T6) successfully fulfilled all three of the binary system criteria. This estimated binary fraction is in good agreement with the observed WR binary fraction in both the LMC \citep{bre1999, neu2012a, mas2014, mas2015a} and SMC\citep{foe2003a}, which again are lower limits, suggesting the mechanism for producing binaries is metallicity independent \citep{she2016}.


\subsection{Stellar Extinction}
\label{sec:ebv}

Interstellar extinction studies of IC10 have previously been attempted using H\,{\sc ii} regions, from which an E(B-V) = 0.83 $\pm$ 0.20 was obtained (Sect.~\ref{sec:neb_ext}). Here we derive new E(B-V) values of individual WR stars based on photometric magnitudes and spectral types. We consider the extinction in two continuum bands, He\,{\sc ii}C at 478nm, and \halpha C at 662nm as follows:

\begin{eqnarray*}
    \label{eq:HeIIextinction1}
    \text{E(He\,{\sc ii}C - \halpha C)} &=& \text{(He\,{\sc ii}C - \halpha C)} - \text{(He\,{\sc ii}C - \halpha C)}_{0} \\
    \label{eq:HeIIextinction1b}
    &=& A_{\text{He\,{\sc ii}C}} - A_{\text{\halpha C}} 
\end{eqnarray*}

The conversion from reddening to E(B-V) was achieved using the Galactic extinction law \citep{sea1979}. Assuming R$_{\lambda}$ = 3.1 A$_{\lambda}$/E(B-V), ratios at wavelengths of 478nm(He\,{\sc ii}C) and 662nm(\halpha C) were determined to be A$_{\text{He\,{\sc ii}C}}$ = 3.7 E(B-V) and A$_{\text{\halpha C}}$ = 2.44 E(B-V) respectively, resulting in:

\begin{equation*}
    E(B-V) = 0.79 [\text{(He\,{\sc ii}C - \halpha C)} - \text{(He\,{\sc ii}C - \halpha C)}_{0}]
	\label{eq:HeIIextinction3}
\end{equation*}


Intrinsic He\,{\sc ii}C-\halpha C values are dependent on WR subtype, therefore for each class, the He\,{\sc ii}C and \halpha C magnitudes were determined from model spectra, free from extinction, to find the intrinsic colour. Where possible we used LMC template WR stars, however there are no late-type WC stars within the LMC so for this case we used a model for a Milky Way WC star. For cases where the WR spectra is dominated by OB stars we use Starburst99 population synthesis models at an age of 5Myr, since this is the typical age of stellar clusters hosting WR stars \citep{lei1999}. Table~\ref{tab:intrinsiccol} lists these results along with their associated model references. 

Where possible we have attempted to derive individual extinction values tailored to each star. A comparison between the nebular and photometrically derived extinctions for T5 (HL45) gives E(B-V)=1.13 $\pm$ 0.04 and E(B-V)=1.15 $\pm$ 0.03 respectively, showing the two methods are in agreement. For the remainder of the sample, where robust magnitude measurements were unavailable, an average extinction value of E(B-V) = 0.92 $\pm$ 0.26 was applied, obtained from both the stellar extinction (E(B-V)=0.95 $\pm$ 0.27), and nebular extinction (E(B-V)=0.83 $\pm$ 0.20) results. The E(B-V) values applied for each star are included in Table~\ref{tab:photdata}.

\begin{table*}
\begin{center}
\caption{Emission line properties of the strongest spectral features found in the confirmed WN and WC stars of IC10. Line luminosities derived using the individual interstellar extinction values outlined in Table~\ref{tab:photdata} and discussed in Sect.~\ref{sec:ebv}, along with a distance of 740 $\pm$ 20 kpc. Mask column refers to the mask the star was observed through. Multiple masks indicate that the star was observed more than once, therefore these spectra were combined to improve signal to noise. For WN stars the strongest lines usually refer to He\,{\sc ii} 4686 and He\,{\sc ii} 5411 emission (with the exception of the WNE/C star M5). For WC stars the strongest features are C\,{\sc iv} 5808 and, due to the broad nature of the emission lines, a C\,{\sc iii} 4650/He\,{\sc ii} 4686 blend. Binary column indicates stars that successfully met two out of the three binary criteria outlined in Sect.~\ref{sec:bf}. Note this only suggests and does not confirm binary status of a star.}
\label{tab:el}
\begin{tabular}{c@{\hspace{1mm}}c@{\hspace{1mm}}c@{\hspace{1mm}}c@{\hspace{-2mm}}c@{\hspace{-2mm}}c@{\hspace{4mm}}c@{\hspace{1mm}}c@{\hspace{-2mm}}		c@{\hspace{-2mm}}c@{\hspace{2mm}}c@{\hspace{1mm}}c@{\hspace{1mm}}c@{\hspace{1mm}}}
\hline
\hline
 & &  \multicolumn{4}{c}{He\,{\sc ii} 4686}                                         & \multicolumn{4}{c}{He\,{\sc ii} 5411} & & Radial &\\
ID & Spectral & FWHM & Log W$_{\lambda}$ & $10^{-17}$ f$_{\lambda}$ & $10^{35}$ L$_{\lambda}$ & FWHM & Log W$_{\lambda}$ & $10^{-17}$ f$_{\lambda}$ & $10^{35}$ L$_{\lambda}$& Mask & Velocity & Binary\\
 & Type & [\ang] & [\ang] & [\ergcms] & [\ergs] & [\ang] & [\ang] & [\ergcms] & [\ergs] & & [kms$^{-1}$] & \\
\hline
T4  & WNE         & 23 $\pm$ 1 & 2.65 $\pm$ 0.02 & 19 $\pm$ 1  &  4.5 $\pm$ 1.6  & 22 $\pm$ 3 & 1.73 $\pm$ 0.05 & 5 $\pm$ 1     & 0.6 $\pm$ 0.2   & 1     & -317 &    \\
R13 & WN5         & 28 $\pm$ 1 & 2.19 $\pm$ 0.02 & 18 $\pm$ 1  &  1.9 $\pm$ 0.8  & 28 $\pm$ 3 & 1.66 $\pm$ 0.04 & 4 $\pm$ 0     & 0.3 $\pm$ 0.1   & 2,3   & -80  &    \\
T5  & WNE         & 32 $\pm$ 0 & 1.08 $\pm$ 0.03 & 20 $\pm$ 3  & 10.9 $\pm$ 3.2  &            &                 &               &                 & 1,2   & -30  & b? \\
R9 & WNE          & 28 $\pm$ 1 & 1.67 $\pm$ 0.02 & 18 $\pm$ 1  &  2.0 $\pm$ 0.2  & 31 $\pm$ 3 & 0.98 $\pm$ 0.04 & 6 $\pm$ 1     & 0.4 $\pm$ 0.1   & 2,3   & -222 &    \\
T6  & WNE         & 27 $\pm$ 2 & 0.81 $\pm$ 0.03 & 46 $\pm$ 3  & 16.6 $\pm$ 2.1  &            &                 &               &                 & 1     & -82  & b? \\
R8 & WN10         & 46 $\pm$ 5 & 1.19 $\pm$ 0.04 & 28 $\pm$ 2  &  2.9 $\pm$ 0.4  &            &                 &               &                 & 4     & -19  & b? \\
M9 & WN3          & 31 $\pm$ 0 & 2.58 $\pm$ 0.01 & 47 $\pm$ 1  &  4.8 $\pm$ 0.9  & 29 $\pm$ 6 & 1.85 $\pm$ 0.07 & 9 $\pm$ 1     & 0.6 $\pm$ 0.1   & 3     & -251 &    \\
R12 & WNE         & 58 $\pm$ 3 & 2.13 $\pm$ 0.03 & 24 $\pm$ 1  & 16.7 $\pm$ 2.5  &            &                 &               &                 & 4     & 259  &    \\
M24 & O2.5 If/WN6 & 21 $\pm$ 1 & 0.88 $\pm$ 0.01 & 163 $\pm$ 4 & 10.4 $\pm$ 1.1  &            &                 &               &                 & 2     & -204 &    \\
R2 & WN7-8        & 18 $\pm$ 0 & 1.62 $\pm$ 0.01 & 31 $\pm$ 1  &  5.9 $\pm$ 0.5  & 17 $\pm$ 4 & 0.79 $\pm$ 0.08 & 6 $\pm$ 1     & 0.6 $\pm$ 0.1   & 2,3   & -117 & b? \\
M17 & WNE + BH    & 20 $\pm$ 4 & 1.01 $\pm$ 0.07 & 8 $\pm$ 1   &  1.2 $\pm$ 0.2  &            &                 &               &                 & 4     &      & var.  \\
M19 & WN4         & 24 $\pm$ 1 & 2.07 $\pm$ 0.01 & 44 $\pm$ 1  &  1.8 $\pm$ 0.3  & 29 $\pm$ 1 & 1.52 $\pm$ 0.01 & 10 $\pm$ 0    & 0.3 $\pm$ 0.04  & 1,3,4 & -260 &    \\
M23 & WN7         & 13 $\pm$ 0 & 1.72 $\pm$ 0.01 & 23 $\pm$ 1  &  3.7 $\pm$ 3.3  & 12 $\pm$ 1 & 0.83 $\pm$ 0.04 & 6 $\pm$ 0     & 0.5 $\pm$ 0.4   & 2     & -286 &    \\
M21 & WN4         & 28 $\pm$ 0 & 2.38 $\pm$ 0.00 & 54 $\pm$ 1  &  7.0 $\pm$ 0.8  & 29 $\pm$ 1 & 1.66 $\pm$ 0.01 & 12 $\pm$ 0    & 0.9 $\pm$ 0.1   & 2,3   & -117 &    \\
\hline
& &  \multicolumn{4}{c}{C\,{\sc iii} 4650/He\,{\sc ii} 4686 blend}                          & \multicolumn{4}{c}{C\,{\sc iv} 5808} & \\
M5 & WNE/C4  & 53 $\pm$ 1 & 2.53 $\pm$ 0.01 & 89 $\pm$ 2  & 21.8 $\pm$ 2.4  & 79 $\pm$ 1 & 3.09 $\pm$ 0.01 & 370 $\pm$ 6   & 38.4 $\pm$ 3.6  & 3,4   & -422 &    \\
M1  & WC4-5  & 65 $\pm$ 3 & 1.86 $\pm$ 0.02 &  46 $\pm$ 2 & 7.5 $\pm$ 6.8   & 51 $\pm$ 1 & 1.99 $\pm$ 0.01 & 84 $\pm$ 2    & 6.3 $\pm$ 4.3   & 4     & 239  & b? \\
M2  & WC4    & 69 $\pm$ 1 & 2.17 $\pm$ 0.01 & 138 $\pm$ 2 & 22.5 $\pm$ 20.3 & 63 $\pm$ 1 & 2.50 $\pm$ 0.01 & 322 $\pm$ 5   & 24.0 $\pm$ 16.4 & 2     & -460 &    \\
R6  & WC4    & 73 $\pm$ 2 & 2.29 $\pm$ 0.01 &  45 $\pm$ 1 & 25.9 $\pm$ 3.7  & 75 $\pm$ 1 & 2.41 $\pm$ 0.01 & 95 $\pm$ 2    & 18.4 $\pm$ 2.1  & 2,4   & -158 & b? \\
R5  & WC4-5  & 74 $\pm$ 1 & 2.16 $\pm$ 0.01 &  64 $\pm$ 1 & 11.3 $\pm$ 1.6  & 72 $\pm$ 5 & 2.18 $\pm$ 0.03 & 99 $\pm$ 6    & 7.9 $\pm$ 1.0   & 2,4   & -183 & b? \\
M4  & WC4-5  & 67 $\pm$ 1 & 1.65 $\pm$ 0.01 & 204 $\pm$ 3 & 32.9 $\pm$ 2.4  & 48 $\pm$ 1 & 1.53 $\pm$ 0.01 & 183 $\pm$ 4   & 13.6 $\pm$ 0.9  & 1     & -163 & b? \\
M7  & WC4-5  & 81 $\pm$ 1 & 2.73 $\pm$ 0.01 & 836 $\pm$ 12& 137 $\pm$ 123   & 79 $\pm$ 2 & 2.44 $\pm$ 0.01 & 1002 $\pm$ 27 & 75 $\pm$ 51     & 1     & -205 &    \\
R11 & WC4    & 83 $\pm$ 4 &                 &  69 $\pm$ 4 & 11.3 $\pm$ 10.2 & 82 $\pm$ 1 &                 & 274 $\pm$ 3   & 20.4 $\pm$ 14.0 & 1     & -445 &    \\
M10 & WC7    & 68 $\pm$ 0 & 3.06 $\pm$ 0.00 & 904 $\pm$ 7 & 496 $\pm$ 41.2  & 84 $\pm$ 1 & 3.06 $\pm$ 0.01 & 1128 $\pm$ 15 & 210 $\pm$ 15.8  & 4     & -47  &    \\
M12 & WC4    & 53 $\pm$ 2 & 1.96 $\pm$ 0.02 &  40 $\pm$ 2 & 80.9 $\pm$ 10.9 & 51 $\pm$ 1 & 1.87 $\pm$ 0.01 & 99 $\pm$ 2    & 49.1 $\pm$ 5.2  & 3     & 10   & b? \\
R10 & WC4    & 61 $\pm$ 2 & 3.39 $\pm$ 0.01 & 84 $\pm$ 2  & 13.7 $\pm$ 12.4 & 52 $\pm$ 1 & 3.13 $\pm$ 0.00 & 118 $\pm$ 1   & 8.8 $\pm$ 6.0   & 1     & -279 &    \\
M13 & WC5-6  & 59 $\pm$ 1 & 2.07 $\pm$ 0.01 & 161 $\pm$ 3 & 8.3 $\pm$ 0.7   & 48 $\pm$ 1 & 1.99 $\pm$ 0.01 & 163 $\pm$ 2   & 5.1 $\pm$ 0.4   & 4     & -59  & b? \\
M14 & WC5    & 77 $\pm$ 1 & 2.64 $\pm$ 0.00 & 588 $\pm$ 6 & 115 $\pm$ 10.7  & 78 $\pm$ 1 & 2.51 $\pm$ 0.01 & 622 $\pm$ 9   & 53.1 $\pm$ 4.4  & 1,3   & -32  & b? \\
M15 & WC4    & 65 $\pm$ 1 & 3.00 $\pm$ 0.01 & 180 $\pm$ 3 & 7.5 $\pm$ 1.8   & 55 $\pm$ 0 & 2.92 $\pm$ 0.00 & 269 $\pm$ 2   & 7.1 $\pm$ 1.3   & 1,2   & -274 &    \\
M20 & WC5    & 61 $\pm$ 0 & 2.94 $\pm$ 0.00 & 272 $\pm$ 2 & 39.2 $\pm$ 5.4  & 53 $\pm$ 1 & 2.82 $\pm$ 0.00 & 269 $\pm$ 3   & 18.2 $\pm$ 2.0  & 1,3   & -232 &    \\

\hline
\hline
\multicolumn{13}{l}{
  \begin{minipage}{0.95\textwidth}~\\
M17 is a known X-ray binary, involving a WR and black hole orbiting with a 34.93 $\pm$ 0.04 hr period. \citep{pre2007, sil2008}.\\
  \end{minipage}
}\\
\end{tabular}
\end{center}
\end{table*}


\subsection{WR Line Luminosities}
\label{sec:lf}

\begin{table*}
\begin{center}
\caption{Average line luminosities of the He\,{\sc ii} 4686, and C\,{\sc iv} 5808 for the WN and WC/O spectral classes respectively. Both the WN and WC class have been further divided using ionization classifications into early and late subgroups. To avoid confusion, the WNE/C star M5 has not been included in any category. For comparison, the mean line luminosities obtained for the LMC and SMC have been included from \citet{cro2006} data.}
\label{tab:avlinelumin}
\begin{tabular}{c@{\hspace{1.5mm}}c@{\hspace{1.5mm}}c@{\hspace{1.5mm}}c@{\hspace{1.5mm}}c@{\hspace{1.5mm}}c@{\hspace{1.5mm}}c@{\hspace{1.5mm}}c@{\hspace{1.5mm}}}
\hline
\hline
 &  & \multicolumn{2}{c}{IC10} & \multicolumn{2}{c}{LMC} & \multicolumn{2}{c}{SMC} \\
Spectral Type & Emission line & WR & L$_{\lambda}$ $\times 10^{35}$ & WR & L$_{\lambda}$ $\times 10^{35}$ & WR & L$_{\lambda}$ $\times 10^{35}$ \\
 & [\ang] & [\#] & [\ergs] & [\#] & [\ergs] & [\#] & [\ergs] \\
\hline
WN2-5 (WNE) & He\,{\sc ii} 4686 & 10 & 6.73 $\pm$ 6.00 & 45 & 9.27 $\pm$ 8.74 & 6  & 0.62 $\pm$ 0.32 \\
WN6-9 (WNL) & He\,{\sc ii} 4686 & 4  & 5.71 $\pm$ 3.37 & 15 & 13.3 $\pm$ 15.2 & 1  & 7.79            \\
WC4-6 (WCE) & C\,{\sc iv} 5808  & 13 & 23.6 $\pm$ 21.8 & 17 & 32.5 $\pm$ 15.8 &    &                 \\
WC7 (WCL)   & C\,{\sc iii} 5696 & 1  & 87 $\pm$ 8      &    &                 &    &                 \\
WC7 (WCL)   & C\,{\sc iv} 5808  & 1  & 203 $\pm$ 18    &    &                 &    &                 \\
WO          & C\,{\sc iv} 5808  &    &                 & 1  & 10.2            & 1  & 14.9            \\
\hline
\hline
\end{tabular}
\end{center}
\end{table*}

Average WR emission line luminosities are very useful for interpreting extragalactic observations of young star forming regions. Significant WR populations can be found in distant galaxies which have recently undergone a burst of massive star formation, however individual WR stars will be unresolved. To probe the WR content we must rely on integrated WR emission line luminosities from the galaxy, calibrated using nearby resolved populations \citep{sch1998, sid2006}.

\begin{figure*}
	\includegraphics[width=\textwidth]{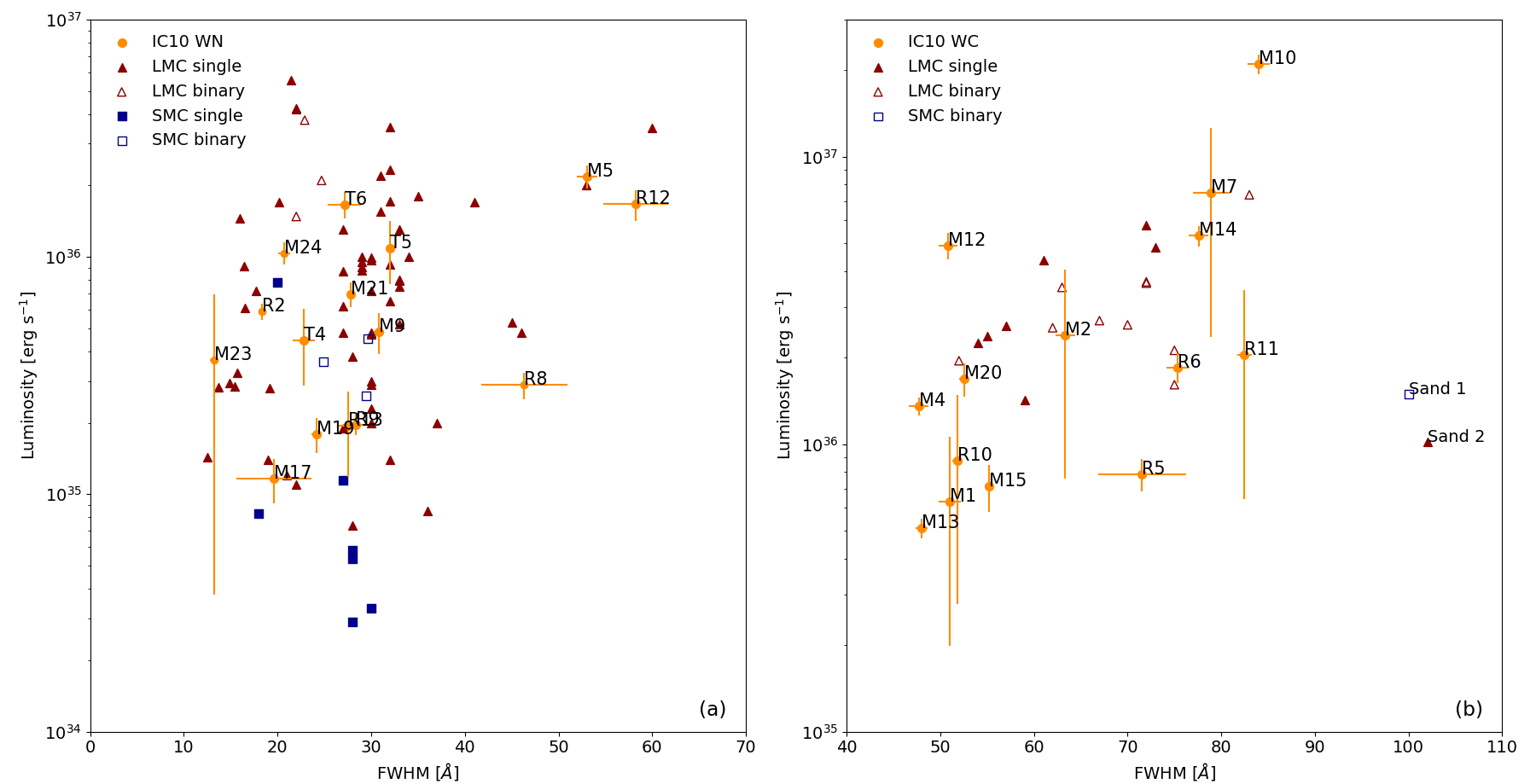}
        \caption{Shows the relationship between FWHM and line luminosity for (a) WN stars using the He\,{\sc ii} 4686 emission line, and (b) WC and WO stars using the C\,{\sc iv} 5808 emission line. The corresponding host galaxy for each WR can be identified from plot symbol and colour using the following; LMC (red triangles), SMC (blue squares), and IC10 (yellow circles). For the Magellanic cloud WR stars, binary and single stars can also be distinguished between by open and filled plot symbols respectively. LMC data taken from \citet{cro2006}, \citet{sch2008}, and \citet{foe2003b}. SMC data from \citet{foe2003a} and \citet{cro2006}.}
    \label{fig:FWHMLumin}
\end{figure*}

Here we present the average WR line luminosities obtained for IC10. We divide the WR population into five categories based on spectral type and determine the luminosities of the strongest emission lines associated with that WR class. The individual stellar extinctions applied for each star are shown in Table~\ref{tab:photdata}. Individual line luminosities can be found in Table~\ref{tab:el} and the average results are summarised in Table~\ref{tab:avlinelumin}. For comparison, we also include Magellanic Cloud WR line luminosity data taken from \citet{cro2006}. Fig.~\ref{fig:FWHMLumin} provides a visual representation for the comparison of individual line luminosities between IC10 and Magellanic Cloud WN, WC and WO stars.

WN stars, using the He\,{\sc ii} 4686 emission line, appear to show conflicting results. For WN2-5 (WNE) stars the average line luminosity of 6.7 $\pm$ 6.0 $\times 10^{35}$ \ergs\ for IC10 is similar to LMC counterparts, whereas for WN6-9 (WNL) stars however, the IC10 average of 5.7 $\pm$ 3.4 $\times 10^{35}$ \ergs\ is somewhat lower than both the Magellanic Clouds. We note however, that the sole late WN star in the SMC is the unusual system HD5980 \citep{koe2014}, and the LMC statistics include the hydrogen rich WN stars in 30 Doradus.

The similarities between IC10 and LMC WN stars can been seen in Fig.~\ref{fig:FWHMLumin}, while SMC WN stars can be seen to have lower luminosities than their LMC and IC10 counterparts.

WCE stars in IC10 and the LMC, which all belong to the WC4 class, have comparable C\,{\sc iv} 5808 emission line averages, however there is no counterpart for the IC10 late WC (WCL) star in either of the Magellanic Clouds. This provides an opportunity to extend the local line luminosity calibrators to include WC7 stars at LMC metallicity. However, our WCL sample consists of only M10. \citet{smi1990} find that Galactic late WC stars have lower C\,{\sc iv} 5808 fluxes than WC4 stars, with an average emission line luminosity of 3.81 $\pm$ 0.46 $\times 10^{35}$ \ergs, therefore suggesting M10 is unusually luminous and may not be typical.

\section{Discussion}
\label{sec:dis}


\subsection{WR Content}
\label{sec:wrcontent}

Table~\ref{tab:localgroup} provides a comparison between IC10 and Magellanic Cloud galactic properties. The size and low metallicity of IC10 has meant that it had previously been considered as a SMC-like galaxy, however when considering the revised metallicity, the presence of WC stars, and average line luminosity comparisons, IC10 bears closer resemblance to the LMC than the SMC.

The total number of spectroscopically confirmed WR stars in IC10 has increased to 29, with a WC/WN ratio of 1.0, putting aside the intermediate WN/C star M5. From Table~\ref{tab:localgroup}, if we take the LMC WR content, which one can consider to be essentially complete \citep{mas2015a}, and scale this to the IC10 SFR of 0.045 \sfr\ we find the expected number of WR stars is $\sim$28, which agrees well with our result and suggests that the IC10 WR survey is also relatively complete.

The WC/WN ratio of IC10 is substantially higher than the LMC value of $\sim$0.2, suggesting an excess of WC stars. Deep imaging reveals a further 6 potential candidates, and if all 6 remaining candidates were confirmed as WN stars, the WC/WN ratio would fall again to 0.7.

Whilst we acknowledge that the method outlined here has limitations regarding the identification of WR stars located in dense local stellar environments, we consider this survey near complete. It should also be noted that a recent survey of the LMC has identified weak emission line WN3 stars, with characteristically faint magnitudes of -2.5 $\pm$ 0.5 mag \citep{mas2014, mas2015a}. The He\,{\sc ii}C sensitivity limit of our survey however is -2.4 mag for a 3$\sigma$ detection, which should be sufficient to identify some of the faint stars belonging to this class. Considering these faint WN3 +abs stars comprise $\sim$ 6\% of the LMC WR population, we are confident a huge hidden population of WR stars does not remain undiscovered within IC10.

\begin{table*}
  \caption{Comparing galactic and individual star forming region properties of Local Group members with similar metallicity to IC10. Distances to the LMC and SMC were taken as 50kpc and 60kpc respectively \citep{wes1990}. 30 Dor. refers to the 30 Doradus star forming region within the LMC, HL 111/106 refers to the central star forming region of IC10. Bracketed values reflect potential changes to the WR number and WC/WN ratio in IC10 should all 6 potential WR candidates be successfully confirmed as WN stars.}
\label{tab:localgroup}
\begin{tabular}{c@{\hspace{1mm}}c@{\hspace{1mm}}c@{\hspace{1mm}}c@{\hspace{1mm}}c@{\hspace{1mm}}c@{\hspace{1mm}}c@{\hspace{1mm}}c@{\hspace{1mm}}c@{\hspace{1mm}}c@{\hspace{1mm}}c@{\hspace{1mm}}c@{\hspace{1mm}}c@{\hspace{1mm}}c@{\hspace{1mm}}}
\hline
\hline
Galaxy & Distance & Log$\frac{O}{H}$+12  &              & $10^{39}$ L$_{H\alpha}$ & SFR   &              & R$_{25}$    & \SFI     &              &  WR  & WC/WN  & Binary      &              \\
       & [kpc]     &                     & \textit{Ref} & [\ergs]              & [\sfr]& \textit{Ref} & [$\arcmin$]& [\sfi]   & \textit{Ref} &  [\#]& [\#]   & Fraction [\%]& \textit{Ref} \\
\hline
LMC          & 50  & 8.37                & a            & 31                   & 0.260 & c            & 646       & 0.0036  & e,f          & 154  & 0.19   & 29          & g,h,i,j  \\
SMC          & 60  & 8.13                & a            & 4.68                 & 0.046 & c            & 316       & 0.0016  & e            & 12   & 0.1    & 42          & k \\
IC10         & 740 & 8.40                & b            & 5.64                 & 0.045 & b            & 6.3       & 0.049   & e            & 29   & 1      & 41:         & b \\
             &     &                     &              &                      &       &              &           &         &              & (35) & (0.7) &             &  \\

30 Dor.      & (LMC)  &                     &              & 14                   & 0.108 & d            & 15*       & 0.722   & d            & 27   & 0.25   & 20          & l \rule{0pt}{4ex} \\ 
HL 111/106   & (IC10) &                     &              & 1.44                 & 0.011 & b            & 0.30$\times$0.46* & 0.559   & b    & 6    & 2      & 67          & b \\
\hline
\hline
\multicolumn{14}{l}{
  \begin{minipage}{2\columnwidth}~\\
    a: \citet{gar1990}; b: this work; c: \citet{ken2008}; d: \citet{ken1995}; e: \citet{cro2009}; f: \citet{bookdev1991} g: \citet{bre1999}; h: \citet{neu2012a}; i: \citet{mas2014}; j: \citet{mas2015a}; k: \citet{foe2003a}; l: \citet{dor2013}; \\
    *R$_{25}$ radius used for all galaxies \citep{bookdev1991} excluding 30 Doradus \citep{ken1995} and the ellipse used for HL 111/106 complex. \\ 
    The colon following the binary fraction highlights the uncertainty in this measurement since a robust method has not been used to confirm potential binary candidates. 
  \end{minipage}
}\\
\end{tabular}
\end{table*}


\subsection{WC/WN Ratio}
\label{sec:wcwn}

The evolution of WR stars from the WN phase to the WC phase is due to mass loss, which is primarily dictated by metal-driven winds, as seen in the clear trend of increasing WC/WN ratio with metallicity in the Local Group \citep{mas2015b, cro2007}. For IC10 however, with an LMC-like metallicity, the high WC/WN ratio remains peculiar, suggesting this relationship also depends on another parameter.
From Table~\ref{tab:localgroup} we see than the SFR in IC10 is unremarkable, however the star formation surface density (\SFI) far exceeds those of the Magellanic Clouds. We therefore consider the \SFI\ as a second parameter in our understanding of WC/WN ratios.

Massive stars are generally formed in clusters \citep{por2010}, for which it is known from the cluster mass function that high mass clusters are rare and low mass clusters are common \citep{whi1999,zha1999}. Increasing the star formation intensity extends the cluster mass function to higher masses, such that the truncation of the upper cluster mass increases for starburst regions with respect to their quiescent star forming counterparts \citep{gie2009}. A second relationship exists between cluster mass and its most massive star, proposed by \citet{wei2006}. Combining these two results allows us to draw the conclusion that regions of more intense star formation are capable of producing higher mass stars, fully sampling the IMF, whereas quiescent star forming regions would exhibit a deficit of high mass stars. This result is significant because the initial mass of the O-star can play a crucial role in the future evolution of the WR star through the WN and WC phases.

When observing LMC WN stars, \citet{hai2014} found that the stellar evolution tracks modelled using the Geneva group stellar evolution models \citep{mey2005} show the majority of WN stars had initial masses within the range of 25-40 \msol. Meanwhile the progenitors of LMC WC stars are likely to have had initial masses greater than 40\msol \citep{cro2002, mey2005}. Therefore larger initial masses are required for single stars to progress to the WC stage.

Comparing IC10 to the LMC, with similar host metallicity environments, we see the global SFR of IC10 is lower but the star formation surface density is an order of magnitude higher (see Table~\ref{tab:localgroup}). The high \SFI\ will extend the stellar mass limit to higher masses, in turn increasing the frequency of higher mass stars. If IC10 has been host to a higher proportion of massive O-stars, the percentage of WR stars capable of achieving the mass-loss rates necessary to reach the WC phase would also increase and the WC/WN ratio would rise to reflect this, as is observed. Indeed, the WC/WN ratio of the dominant star-forming complex of IC10, comprising of HL106/111 \citep{hod1990} is especially high, as summarised in Table~\ref{tab:localgroup}.

Within the Local Group, the closest analogue to the high \SFI\ of IC10 is the 30 Doradus region in the LMC. A census of the WR content of 30 Doradus, within 15\arcmin of R136, implies a ratio of WC/WN = 0.25 \citep{bre1999, dor2013}. However, putting aside main sequence very massive WN5h(a) stars, the WC/WN ratio rises to 0.42. Again, this increased ratio arises from the high \SFI\ of 30 Doradus, leading to an increased frequency of high mass stars in this region and consequently a higher WC/WN ratio. Similarly, a low \SFI\ at high metallicity would produce a low WC/WN ratio. By way of example, the super-solar metallicity galaxy M31 has a relatively low \SFI\ and a modest WC/WN = 0.67 ratio \citep{neu2012b}.



\section{Conclusion}
\label{sec:con}


Our main results can be summarized as follows:

\begin{enumerate}

\item Using deep narrow-band imaging to search for a He\,{\sc ii} 4686 magnitude excess, we present 11 WR candidates in IC10 and spectroscopically confirm 3 of these as WN stars, whilst rejecting 1 as an early M-dwarf and suggesting another is unlikely to be a WR star due to the dispersed nature of the source. The total number of WR stars in IC10 has now been raised from 26 to 29, and the WC/WN ratio has lowered to 1.0. We review previous spectral classifications and suggest updates for 3 previously confirmed WR stars, M15, M23 and M24.

\item An updated SFR measurement of 0.045 $\pm$ 0.023 \sfr\ has been derived from \halpha\ luminosity, an increase from the previous result of 0.031 $\pm$ 0.007 \citep{ken2008}. This updated \halpha\ SFR is intermediate between radio derived SFRs, however we note the radio fluxes have not been corrected to eliminate non-thermal radio sources.  

\item Using nebular emission from the H\,{\sc ii} region HL45, associated with the WR star T5, the oxygen abundance for IC10 has also been updated from 8.26 to 8.40 $\pm$ 0.04, suggesting IC10 has an LMC-like metallicity. Comparison of emission line luminosities also revealed similar results for WNE and WCE stars in IC10 and the LMC, emphasising the similarities, however the WNL and WCL results were less consistent, most likely due to the small number of stars in these categories.

\item The WC/WN ratio observed for IC10 remains peculiar, despite the potential addition of our 6 new unconfirmed candidates. We propose the most likely explanation is due to the high star formation surface density of the galaxy, which extends the cluster upper stellar mass limit to higher values. Assuming the WC initial mass limit is higher than for WN stars, the higher WC/WN ratio observed in IC10 would be expected as a result of the higher star formation intensity observed in this galaxy.
  
\end{enumerate}


\section*{Acknowledgements}

KT and IA would like to thank STFC for financial support. Also thanks to Laurent Drissen (co-PI for original Gemini dataset) and our anonymous referee for their careful review and helpful comments.
Based on observations obtained at the Gemini Observatory acquired through the Gemini Observatory Archive, which is operated by the Association of Universities for Research in Astronomy, Inc., under a cooperative agreement with the NSF on behalf of the Gemini partnership: the National Science Foundation (United States), the National Research Council (Canada), CONICYT (Chile), Ministerio de Ciencia, Tecnolog\'{i}a e Innovaci\'{o}n Productiva (Argentina), and Minist\'{e}rio da Ci\^{e}ncia, Tecnologia e Inova\c{c}\~{a}o (Brazil).

The Pan-STARRS1 Surveys (PS1) have been made possible through contributions of the Institute for Astronomy, the University of Hawaii, the Pan-STARRS Project Office, the Max-Planck Society and its participating institutes, the Max Planck Institute for Astronomy, Heidelberg and the Max Planck Institute for Extraterrestrial Physics, Garching, The Johns Hopkins University, Durham University, the University of Edinburgh, Queen's University Belfast, the Harvard-Smithsonian Center for Astrophysics, the Las Cumbres Observatory Global Telescope Network Incorporated, the National Central University of Taiwan, the Space Telescope Science Institute, the National Aeronautics and Space Administration under Grant No. NNX08AR22G issued through the Planetary Science Division of the NASA Science Mission Directorate, the National Science Foundation under Grant No. AST-1238877, the University of Maryland, and Eotvos Lorand University (ELTE).


\bibliographystyle{mnras}


\appendix

\section{Candidate WR Stars}
\label{appen:potentialWR}

\begin{table*}
\begin{center}
\caption{He\,{\sc ii} 4686 emission candidates identified from GMOS He\,{\sc ii} and He\,{\sc ii}C filter images, using blinking and subtraction techniques to search for He\,{\sc ii} excesses. Candidates listed in increasing RA order.}
\label{tab:potentialWR}
\begin{tabular}{c@{\hspace{1.5mm}}c@{\hspace{1.5mm}}c@{\hspace{1.5mm}}c@{\hspace{1.5mm}}c@{\hspace{1.5mm}}c@{\hspace{1.5mm}}c@{\hspace{1.5mm}}c@{\hspace{1.5mm}}c@{\hspace{1.5mm}}c@{\hspace{1.5mm}}}
\hline
\hline
ID & RA & Dec & He\,{\sc ii} & He\,{\sc ii}C & \halpha & \halpha C & g & $\Delta$ He\,{\sc ii}-He\,{\sc ii}C & Status \\
 & \multicolumn{2}{c}{[J2000]} & [mag] & [mag] & [mag] & [mag] & [mag] & [mag] &  \\
\hline
T1  & 00:20:04.54 & 59:18:05.4 & 24.119 $\pm$ 0.030 & 24.983 $\pm$ 0.035 & 22.312 $\pm$ 0.060 & 23.621 $\pm$ 0.117 & 24.213 $\pm$ 0.091 & -0.86 $\pm$ 0.05 &  \\
T2  & 00:20:05.60 & 59:19:45.7 & 24.312 $\pm$ 0.036 & 24.633 $\pm$ 0.053 &                    &                    & 24.476 $\pm$ 0.168 & -0.32 $\pm$ 0.06 &  \\
T3  & 00:20:06.99 & 59:17:47.1 & 24.211 $\pm$ 0.028 & 25.521 $\pm$ 0.053 & 22.406 $\pm$ 0.055 &  & 24.489 $\pm$ 0.162 & -1.31 $\pm$ 0.06 & \\
T4  & 00:20:14.47 & 59:18:49.9 & 23.249 $\pm$ 0.016 & 24.612 $\pm$ 0.034 & 22.765 $\pm$ 0.123 & 23.377 $\pm$ 0.121 & 23.958 $\pm$ 0.093 & -1.36 $\pm$ 0.04 & WNE \\
T5  & 00:20:17.43 & 59:18:39.2 & 21.054 $\pm$ 0.044 & 20.918 $\pm$ 0.054 &                    & 19.532 $\pm$ 0.069 &                    &  0.14 $\pm$ 0.07 & WNE\\
T6  & 00:20:20.34 & 59:18:37.3 & 19.939 $\pm$ 0.019 & 19.801 $\pm$ 0.027 & 18.125 $\pm$ 0.022 & 18.561 $\pm$ 0.018 & 19.222 $\pm$ 0.018 &  0.14 $\pm$ 0.03 & WNE\\
T7  & 00:20:23.35 & 59:17:31.2 & 23.098 $\pm$ 0.027 & 23.530 $\pm$ 0.065 &                    & 22.205 $\pm$ 0.064 & 22.231 $\pm$ 0.037 & -0.43 $\pm$ 0.07 &  \\
T8  & 00:20:27.70 & 59:19:15.1 & 24.671 $\pm$ 0.056 & 25.487 $\pm$ 0.059 &  &  &  & -0.82 $\pm$ 0.08 & extended \\
T9  & 00:20:32.74 & 59:15:46.4 & 22.570 $\pm$ 0.011 & 22.537 $\pm$ 0.008 & 19.214 $\pm$ 0.011 & 19.931 $\pm$ 0.012 & 21.635 $\pm$ 0.011 &  0.03 $\pm$ 0.01 & non WR \\
T10 & 00:20:32.98 & 59:18:24.1 & 24.373 $\pm$ 0.042 & 25.111 $\pm$ 0.077 &                    & 23.467 $\pm$ 0.089 &                    & -0.74 $\pm$ 0.09 &  \\
T11 & 00:20:35.90 & 59:18:49.8 & 23.055 $\pm$ 0.019 & 23.428 $\pm$ 0.022 & 21.894 $\pm$ 0.060 & 22.888 $\pm$ 0.088 &                    & -0.37 $\pm$ 0.03 &  \\
\hline
\hline
\end{tabular}
\end{center}
\end{table*}

\section{Balmer Emission Line Strengths}
\label{appen:ew}

\begin{table}
  \caption{Balmer emission line equivalent widths for the H\,{\sc ii} regions outlined in Sect.~\ref{sec:neb_ext}. As before, HL\# refer to H\,{\sc ii} regions outlined by \citet{hod1990} and H\,{\sc ii}\# refer to candidate H\,{\sc ii} regions suggested by \citet{roy2001}.}
  \label{tab:nebularew} \begin{tabular}{c@{\hspace{1.5mm}}c@{\hspace{1.5mm}}c@{\hspace{1.5mm}}c@{\hspace{1.5mm}}c@{\hspace{1.5mm}}}
\hline
\hline
Nebular & Mask & Log W$_{H\gamma}$ & Log W$_{H\beta}$ & Log W$_{H\alpha}$  \\
Region  &      & [\ang]        & [\ang]       & [\ang]         \\
\hline
HL 6          &2&                   & 0.469 $\pm$ 0.067 & 1.205 $\pm$ 0.006 \\
HL 10         &2&                   & 1.043 $\pm$ 0.034 & 1.759 $\pm$ 0.004 \\
HL 20         &3&                   & 0.960 $\pm$ 0.020 & 1.773 $\pm$ 0.004 \\
HL 22         &4&                   & 1.988 $\pm$ 0.004 & 2.685 $\pm$ 0.003 \\
HL 45         &3&                   & 1.940 $\pm$ 0.010 & >3.00 \\
HL 45         &1& 1.613 $\pm$ 0.004 & 2.176 $\pm$ 0.006 &                   \\
H\,{\sc ii} 04&4&                   & 0.952 $\pm$ 0.014 & 1.661 $\pm$ 0.023 \\
H\,{\sc ii} 07&3&                   & 2.654 $\pm$ 0.006 & 3.248 $\pm$ 0.002 \\
H\,{\sc ii} 07&4&                   & 2.892 $\pm$ 0.012 & 3.521 $\pm$ 0.002 \\
H\,{\sc ii} 08&3&                   & 2.565 $\pm$ 0.002 & 3.447 $\pm$ 0.002 \\
H\,{\sc ii} 08&4& 2.407 $\pm$ 0.006 & 2.695 $\pm$ 0.006 &                \\
H\,{\sc ii} 11&3&                   & >3.00             &                \\
\hline
\hline
\multicolumn{5}{l}{
  \begin{minipage}{0.8\columnwidth}~\\
     RA and DEC (J2000) co-ordinates for \citet{roy2001} H\,{\sc ii} regions as follows: H\,{\sc ii} 04 (00:20:15.48, +59:18:40.6) H\,{\sc ii} 07 (00:20:18.51, +59:17:40.4) H\,{\sc ii} 08 (00:20:24.41, +59:16:55.2) H\,{\sc ii} 11 (00:20:19.36, +59:18:02.9) \\
  \end{minipage}
}\\
  \end{tabular}
\end{table}

\section{Model WR Intrinsic Colours}
\label{appen:intrinsiccol}

\begin{table}
\caption{The (He\,{\sc ii}C-\halpha C)$_{0}$ intrinsic colour of WR stars for different ionization subclasses, determined from the spectral model indicated by the reference column.}
\label{tab:intrinsiccol}
\begin{tabular}{c@{\hspace{3mm}}c@{\hspace{3mm}}c@{\hspace{3mm}}c@{\hspace{3mm}}}
\hline
\hline
Spectral type & Template Star & (He\,{\sc ii}C-\halpha C)$_{0}$ & Ref \\
\hline
WN3-4  & LMC-AB9    & -0.06 & a \\
WN6    & HD 38282   & 0.03  & a \\
WN7    & HDE 269883 & -0.01 & a \\
WN8    & LMC-AB11   &  0.00 & a \\
WN10   & BE 294     & 0.12  & a \\
WC4-5  & HD 37026   & -0.05 & b \\
WC6    & HD 97809   & -0.06 & c \\
WC7    & HD 156385  & 0.00  & d \\
WR +OB &            & -0.21 & e  \\
\hline
\hline
\multicolumn{4}{l}{
  \begin{minipage}{\columnwidth}~\\
a: \citet{dor2013}; b: \citet{cro2002}; c: \citet{sma2001}; d: \citet{des2000}; e: \citet{lei1999} \\
  \end{minipage}
}\\
\end{tabular}
\end{table}

\bsp	
\label{lastpage}
\end{document}